\documentclass[10pt,conference]{IEEEtran}
\usepackage{multirow}
\usepackage{xurl}
\usepackage{rotating}
\usepackage{booktabs} % For formal tables
\usepackage{amsmath}
\usepackage{graphicx}
\usepackage[belowskip=-12pt]{caption}
\usepackage{subfigure}
\usepackage{subfloat}
\usepackage{float}
\usepackage{rotating}
\usepackage{tabularx}
\usepackage{booktabs}
\usepackage{amssymb}
\usepackage{wrapfig}
\usepackage{listings}
\usepackage{color, colortbl}
\usepackage{balance} % to balance the bibliography
\usepackage{lscape}
\usepackage{hyperref}
\usepackage{xspace}
\usepackage{framed}
\usepackage{syntax}
\usepackage{enumitem}
\usepackage{siunitx, makecell}
\usepackage{array}

\usepackage{soul}
\usepackage{flushend}
\usepackage[tikz]{bclogo}
\usepackage{makecell}
\definecolor{codegreen}{rgb}{0,0.6,0}
\definecolor{codegray}{rgb}{0.5,0.5,0.5}
\definecolor{codepurple}{rgb}{0.58,0,0.82}
\definecolor{backcolour}{rgb}{0.95,0.95,0.92}
\definecolor{redColor}{RGB}{255,0,0}
\definecolor{Gray}{gray}{0.1}
\lstdefinestyle{mystyle}{
	backgroundcolor=\color{backcolour},   
	commentstyle=\color{codegreen},
	keywordstyle=\color{magenta},
	numberstyle=\tiny\color{codegray},
	stringstyle=\color{codepurple},
	basicstyle=\scriptsize,
	breakatwhitespace=false,         
	breaklines=true,                 
	captionpos=b,                    
	keepspaces=true,                 
	numbers=left,                    
	numbersep=5pt,                  
	showspaces=false,                
	showstringspaces=false,
	showtabs=false,                  
	tabsize=2
}

\lstdefinelanguage{Pythonna}{%
	language     = python,
	breaklines = true,
	morekeywords = {to_categorical, flow_from_directory,pad_sequences,load_image},
	morecomment=[f][\color{diffstart}]{@@},
	morecomment=[f][\color{codegreen}]{+\ },
	morecomment=[f][\color{redColor}]{-\ }
}

\lstdefinestyle{customc}{
	belowcaptionskip=1\baselineskip,
	breaklines=false,
	frame= single,
	breaklines = true,
	xleftmargin=\parindent,
	language= Pythonna,
	showstringspaces=false,
	keywordstyle=\bfseries\color{green!40!black},
	commentstyle=\itshape\color{purple!40!black},
	identifierstyle=\color{blue},
	stringstyle=\color{codegreen},
	backgroundcolor=\color{gray!4}
}

\graphicspath{{./figures/}}

\newcommand{\rnn}{\text{RNN}\xspace}

\newcommand{\relu}{\textit{ReLU}\xspace}
\newcommand{\rolled}{\textit{rolled}\xspace}
\newcommand{\unrolled}{\textit{unrolled}\xspace}
\newcommand{\one}{\textit{one}\xspace}
\newcommand{\many}{\textit{many}\xspace}

\newcommand{\ts}{\textit{timestep}\xspace}
\newcommand{\tss}{\textit{timesteps}\xspace}

\newcommand{\figref}[1]{Fig.~\ref{#1}}

\newcommand{\etal}{{\em et al.}\xspace}

\newcounter{rqs}
\stepcounter{rqs}

\newcounter{NumObservations}
\stepcounter{NumObservations}
\definecolor{shadecolor}{rgb}{.9,.9,.9}

\lstdefinestyle{mystyle}{
	backgroundcolor=\color{backcolour},   
	commentstyle=\color{codegreen},
	keywordstyle=\color{magenta},
	numberstyle=\tiny\color{codegray},
	stringstyle=\color{codepurple},
	basicstyle=\scriptsize,
	breakatwhitespace=false,         
	breaklines=true,                 
	captionpos=b,                    
	keepspaces=true,                 
	numbers=left,                    
	numbersep=5pt,                  
	showspaces=false,                
	showstringspaces=false,
	showtabs=false,                  
	tabsize=2
}
 \usepackage{colortbl}
 \usepackage{algorithm}
 \usepackage[noend]{algpseudocode}
 \usepackage{xspace}
 \usepackage{booktabs, multirow} % for borders and merged ranges
 \usepackage{soul}% for underlines
 \usepackage{float}
 \usepackage{placeins}
 \usepackage[numbers]{natbib}
\AtBeginDocument{%
	\providecommand\BibTeX{{%
			\normalfont B\kern-0.5em{\scshape i\kern-0.25em b}\kern-0.8em\TeX}}}

\graphicspath{{./figures/}}
\begin{document}

%\title{On the Criteria to Decompose Deep Learning Model into Modules}
%\title{Cool Modules for DL Models}
% \title{Decomposing Recurrent Neural Networks into Modules}
% COMMENT: Being more explicit about benefits/motivation
\title{Decomposing a Recurrent Neural Network into Modules for Enabling Reusability and Replacement}

% \author{\IEEEauthorblockN{Sayem Mohammad Imtiaz}
% \IEEEauthorblockA{\textit{Department of Computer Science} \\
% \textit{Iowa State University}\\
% Ames, IA, USA \\
% sayem@iastate.edu}
% \and
% \IEEEauthorblockN{Fraol Batole}
% \IEEEauthorblockA{\textit{Department of Computer Science} \\
% \textit{Iowa State University}\\
% Ames, IA, USA \\
% fraol@iastate.edu}
% \and
% \IEEEauthorblockN{Astha Singh}
% \IEEEauthorblockA{\textit{Department of Computer Science} \\
% \textit{Iowa State University}\\
% Ames, IA, USA \\
% asthas@iastate.edu}
% \and
% \IEEEauthorblockN{Rangeet Pan\IEEEauthorrefmark{1}}
% \IEEEauthorblockA{\textit{Research Staff Member} \\
% \textit{IBM Research }\\
% Yorktown Heights, NY \\
% rangeet.pan@ibm.com}
% \and
% \IEEEauthorblockN{Breno Dantas Cruz}
% \IEEEauthorblockA{\textit{Department of Computer Science} \\
% \textit{Iowa State University}\\
% Ames, IA, USA \\
% bdantasc@iastate.edu}
% \and
% \IEEEauthorblockN{Hridesh Rajan}
% \IEEEauthorblockA{\textit{Department of Computer Science} \\
% \textit{Iowa State University}\\
% Ames, IA, USA \\
% hridesh@iastate.edu}
% }

\author{%
  \IEEEauthorblockN{%
    Sayem Mohammad Imtiaz\IEEEauthorrefmark{1},
    Fraol Batole\IEEEauthorrefmark{1},
    Astha Singh\IEEEauthorrefmark{1},
    Rangeet Pan\IEEEauthorrefmark{2}\textsuperscript{\textsection},
    Breno Dantas Cruz\IEEEauthorrefmark{1}, and
    Hridesh Rajan\IEEEauthorrefmark{1}%
  }%
  \IEEEauthorblockA{\IEEEauthorrefmark{1} \textit{Department of Computer Science, Iowa State University}, Ames, IA, USA}%
  \IEEEauthorblockA{\IEEEauthorrefmark{1}\{sayem, fraol, asthas, bdantasc, hridesh\}@iastate.edu}%
  \IEEEauthorblockA{\IEEEauthorrefmark{2} \textit{IBM Research}, Yorktown Heights, NY, USA, rangeet.pan@ibm.com}%

% \IEEEauthorblockA{\IEEEauthorrefmark{1} \{sayem,fraol,asthas,bdantasc,hridesh\}@iastate.edu}, rangeet.pan@ibm.com%
}

\maketitle
\begingroup\renewcommand\thefootnote{\textsection}
\footnotetext{At the time this work was completed, Rangeet Pan was a graduate student at Iowa
State University}
\endgroup
\thispagestyle{plain}
\pagestyle{plain}
\begin{abstract}
	Can we take a recurrent neural network (RNN) trained to translate between languages
and augment it to support a new natural language without retraining the model from scratch?
Can we fix the faulty behavior of the RNN by replacing portions associated with the 
faulty behavior?
Recent works on decomposing a fully connected neural network (FCNN) and convolutional
neural network (CNN) into modules have shown the value of engineering deep models 
in this manner, which is standard in traditional SE but foreign for deep learning models. 
However, prior works focus on the image-based multi-class classification problems 
and cannot be applied to RNN due to (a) different layer structures, (b) loop structures, (c) different types of input-output architectures, and (d) usage of both non-linear and logistic activation functions.
In this work, we propose the first approach to decompose an RNN into modules. We study different types of RNNs, i.e., Vanilla, LSTM, and GRU.
Further, we show how such RNN modules can be reused and replaced in various scenarios. 
We evaluate our approach against 5 canonical datasets (i.e., Math QA, Brown Corpus, 
Wiki-toxicity, Clinc OOS, and Tatoeba) and 4 model variants for each dataset. 
We found that decomposing a trained model has a small cost (Accuracy: \textbf{-0.6\%}, 
BLEU score: \textbf{+0.10\%}). 
Also, the decomposed modules can be reused and replaced without needing to retrain.

\end{abstract}

\section{Introduction}
\label{sec:intro}

Recurrent neural networks (RNNs), like fully-connected neural networks (FCNN) and convolutional neural networks (CNN), are a class of deep learning (DL) algorithms that are critical for important problems such as text classification. Depending on their architecture, they are further classified into vanilla RNN, LSTM (Long Short Term Memory), or GRU (Gated Recurrent Unit). 
To build such models, the most common way is by training from scratch. 
Otherwise, developers can also use transfer learning~\cite{pratt1991direct, zhang2020adapnet} to reuse a model by retraining its last few layers. Such types of model reuse are coarse-grained, relying on the original model's entire structure. %Moreover, those approaches require additional data, which is often expensive and hard to collect~\cite{cost}.
In contrast, we propose to decompose a trained \rnn model to enable fine-grained reuse without needing to retrain.  
\begin{table}[htbp]
  \centering
  \scriptsize
%   \vspace{5pt}
  \caption{Comparison with the existing works}
%   \vspace{-5pt}
   \setlength\tabcolsep{1.2pt}
   % \resizebox{0.47\textwidth}{!}{%
    \begin{tabular}{|c|l|c|c|c|}
    \hline
    \multicolumn{2}{|c|}{\textbf{Functionality}} & \textbf{FCNN-D} & \textbf{CNN-D} & \textbf{Our Work} \\
    \hline
    \hline
    \multirow{2}{*}{\textbf{Input}} & Image & \textcolor{teal}{\textbf{\checkmark}}     & \textcolor{teal}{\textbf{\checkmark}}     & \textcolor{teal}{\textbf{\checkmark}} \\
\cline{2-5}          & Sequential Data (e.g., Text) & \textcolor{red}{\textbf{X}}     & \textcolor{red}{\textbf{X}}     & \textcolor{teal}{\textbf{\checkmark}} \\
% \cline{2-5}          & Natural Language & \textcolor{red}{\textbf{X}}     & \textcolor{red}{\textbf{X}}     & \textcolor{teal}{\textbf{\checkmark}} \\
    \hline
    \multicolumn{1}{|c|}{\multirow{6}{*}{\textbf{\makecell{Model \\ Properties}}}} & \makecell[l]{Models with more than one output} & \textcolor{red}{\textbf{X}}     & \textcolor{red}{\textbf{X}}     & \textcolor{teal}{\textbf{\checkmark}} \\
\cline{2-5}          & \makecell[l]{Models classify into one of the labels} & \textcolor{teal}{\textbf{\checkmark}}     & \textcolor{teal}{\textbf{\checkmark}}     & \textcolor{teal}{\textbf{\checkmark}} \\
\cline{2-5}          & Models with loops & \textcolor{red}{\textbf{X}}     & \textcolor{red}{\textbf{X}}     & \textcolor{teal}{\textbf{\checkmark}}\\
\cline{2-5}          & \makecell[l]{Models with more than one input} & \textcolor{red}{\textbf{X}}    & \textcolor{red}{\textbf{X}}     & \textcolor{teal}{\textbf{\checkmark}} \\
\cline{2-5}          & Shared weight and bias & \textcolor{red}{\textbf{X}}     & \textcolor{teal}{\textbf{\checkmark}}     & \textcolor{teal}{\textbf{\checkmark}} \\
\cline{2-5}          & Gated layer architecture & \textcolor{red}{\textbf{X}}     & \textcolor{red}{\textbf{X}}    & \textcolor{teal}{\textbf{\checkmark}}  \\
\cline{2-5}          & Non-linear activation function & \textcolor{teal}{\textbf{\checkmark}}      & \textcolor{teal}{\textbf{\checkmark}}      & \textcolor{teal}{\textbf{\checkmark}}  \\
\cline{2-5}          & Logistic activation function & \textcolor{red}{\textbf{X}}     & \textcolor{red}{\textbf{X}}     & \textcolor{teal}{\textbf{\checkmark}}  \\
    % \hline
    % \textbf{Model Parameter} & Supported activation functions & ReLU  & ReLU  & ReLU\\
    \hline
    \end{tabular}%
    % }
  \label{tb:compare}\\
%   \vspace{1pt}
  * FCNN-D: Fully Connected Neural Network decomposition approach~\cite{pan2020decomposing}, CNN-D: Convolutional Neural Network decomposition approach~\cite{pan2022decomposing}
\end{table}%

The term `{\em modules}' has also appeared in the AI/ML community; however, it serves a different purpose~\cite{andreas2016neural, hu2017learning, hinton2000learning, sabour2017dynamic, ghazi2019recursive}. 
They aim to add external memory capability to the DL models~\cite{ghazi2019recursive}. 
To illustrate, Ghazi~\etal presents an example of a room and objects within~\cite{ghazi2019recursive}. A DL model is excellent in answering immediate questions such as "Is there a cat in the room"? However, suppose a person often visits a room over many years and later ponders an indirect question, "How often was there a cat?". In that case, a DL model is incapable of answering it. As a remedy, this line of work proposes to build a deep modular network consisting of many independent neural networks~\cite{sabour2017dynamic,ghazi2019recursive,andreas2016neural}. Essentially, they view a module as a function created in advance. Once a problem is given, they dynamically instantiate a composition of the modules to answer such questions. In all these cases, the end result is still akin to a monolithic model tasked with solving a particular problem. On the contrary, we aim to decompose a trained \rnn model to enable the benefit of software decomposition.

Along this line, recent work has proposed an approach to decompose FCNN and CNN models 
into modules and enable their reuse~\cite{pan2020decomposing, pan2022decomposing}.
\begin{figure*}[]
	\centering
	\includegraphics[width=1.0\linewidth,trim=0cm 0cm 0cm 0cm]{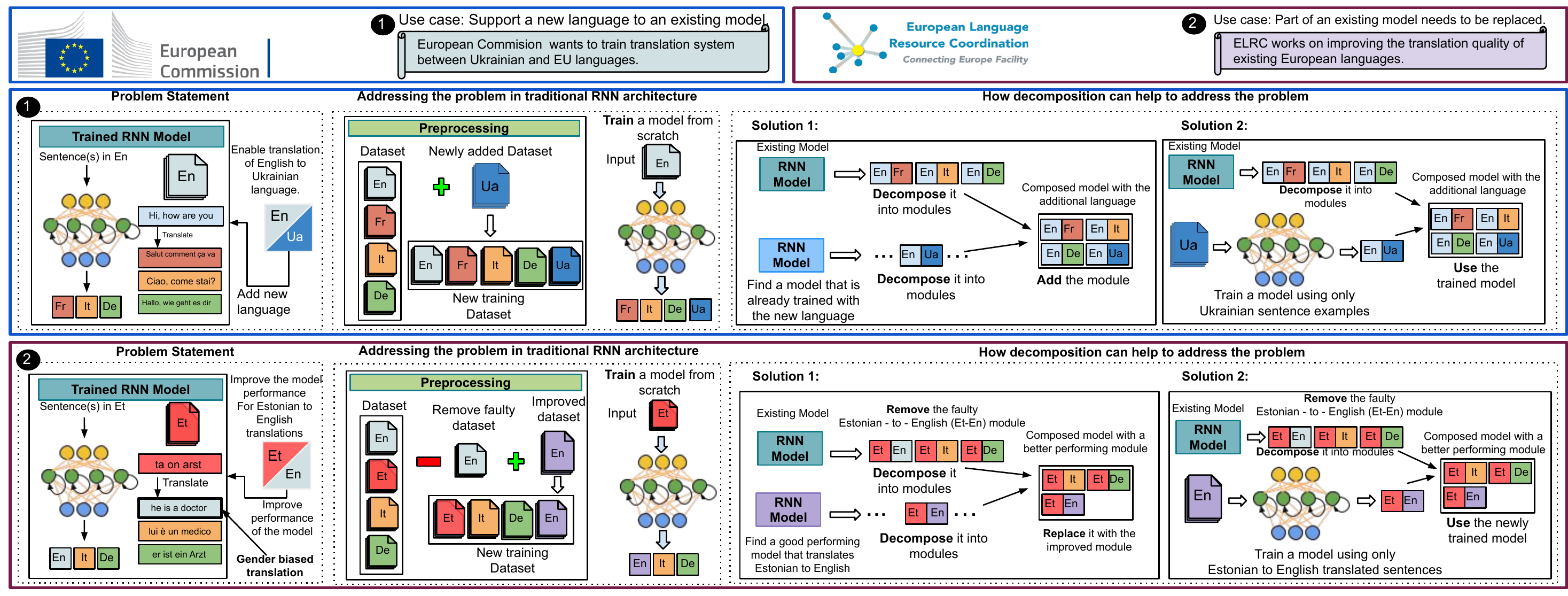}
	\vspace{-20pt}
	\caption{Motivating Examples. De: German, En: English, Et: Estonian, Fr: French, It: Italian, Ua: Ukrainian Language.}
	\label{fig:motivation}
\end{figure*}
However, those approaches cannot be applied to RNNs due to the challenges listed in Table~\ref{tb:compare}. For instance, RNNs, particularly LSTM and GRU, incorporate activation functions such as Tanh and Sigmoid in their internal architecture, which are unsupported by prior works. RNNs have five types of architecture depending on the network's input-output. Moreover, RNN includes a loop structure to process sequence data effectively. These differences render prior approaches inapplicable to RNNs. 
 
Therefore, in this work, we ask: can we identify the RNN model parts responsible for each task and decompose them into modules? Doing so would allow us to (a) build a new problem or (b) replace existing model functionality. In both cases, this type of reuse removes the need for additional data and training. To that end, we propose an approach for decomposing RNNs into modules. 
A key innovation in our work is handling the loop in \rnn in both time insensitive and sensitive manner. 
Inspired by prior works on understanding loops in SE~\cite{dongarra1979unrolling, weiss1987study, davidson1995improving}, we propose to identify nodes and edges responsible for each output class - (a) over all the iterations of the loop (\rolled) and (b) individually, for each iteration in the loop
(\unrolled). \textit{Unrolled}-variant is aware of the time dimension of the model, while \rolled is not. Second, to handle different RNN architectures, we identify the concern (i.e., parts of the network responsible for classifying an output label) and untangle each output timestep at a time. In prior works, each concern is identified and untangled separately; this does not apply to models that produce many outputs. Third, we support models built using logistic activation functions, i.e., Tanh, Sigmoid, etc., 
which are commonly used in RNN~\cite{cho2014properties,hochreiter1997long,lipton2015critical}. 
In addition, we propose a decomposition approach assuming \relu activation as well.

To evaluate our approach, we apply it to five different input-output (I/O) architecture types.
Moreover, we evaluate different RNN-variants (LSTM, GRU, Vanilla) for each architecture. To that end, we utilize Math QA~\cite{amini2019mathqa}, Brown Corpus~\cite{francis1967computational}, Wiki-toxicity~\cite{wulczyn2017ex}, Clinc OOS~\cite{larsonetal2019evaluation}, and Tatoeba~\cite{TIEDEMANN12.463} datasets for training models in different setups, and decompose them. In total, our benchmark consists of 60 models, i.e., 4 (\# models) * 3 (\# RNN-variants) * 5 (\# I/O architectures). In this work, we use the terms ``RNN'' or ``recurrent model'' interchangeably to refer to three RNN-variants collectively.

\textit{Key Results:}
To evaluate our approach, first, we measure the cost of decomposition by comparing the accuracy of the model composed using decomposed modules and the monolithic model from which the modules are decomposed. We found that the loss of accuracy is trivial (Avg.:-0.6\%, median: -0.24\%). For language translation models, there is a slight gain in performance (Avg.: +0.10\%, median: +0.01\%), measured in
BLEU score~\cite{papineni2002bleu}. We also find that the decomposition of models producing more than one output must be time-sensitive or aware of what output appears at what time. 
Second, we evaluated our approach to reusing and replacing the decomposed modules to build various new problems. We compared the accuracy of the models composed using decomposed modules with monolithic ones, trained from scratch. When reusing and replacing, we found that the performance change is 
(accuracy: -2.38\%, BLEU: +4.40\%) and (accuracy: -7.16\%, BLEU: +0.98\%), on average, respectively. \textbf{All the results and code for replication is available here~\cite{rnnrep}.}

The key contributions of this work are as follows:
\begin{itemize}
	\item We propose an approach to decompose an RNN model. 
	\item Our proposed approach is applicable for all five types of I/O architectures.
	\item We propose two variants to support loops in RNN.
	\item Our approach supports both logistic and \relu activation and all 3 RNNs, i.e., Vanilla, LSTM, and GRU. 
	\item We show that our approach can reuse and replace the modules to build new problems without retraining.

\end{itemize}

% We provide an example in \S\ref{sec:motivation} to illustrate how our approach can help developers solve real-world problems. We discuss the closest related works in \S\ref{sec: related}. Then, we discuss the approach in \S\ref{sec:approach}. In \S\ref{sec:evaluation}, we evaluate our approach against three research questions, and provide the concluding remarks in~\S\ref{sec:conclusion}.

\section{Motivating Examples}
\label{sec:motivation}

In this section, we show two examples of using RNN models and how decomposing them into modules could help. The RNN models are used for multilingual language translation. Note that it is preferably performed in a multilingual setup to improve the overall performance of the translation task~\cite{fan2021beyond,chollet2021deep}. \figref{fig:motivation} shows two examples in which decomposition can assist developers when building RNN models for translation.

\textbf{Adding a new language to the European Union system.}
\begin{figure*}[]
	%\vspace{32pt}
	\centering
	\includegraphics[width=\linewidth, trim=0cm 0cm 0cm 0cm]{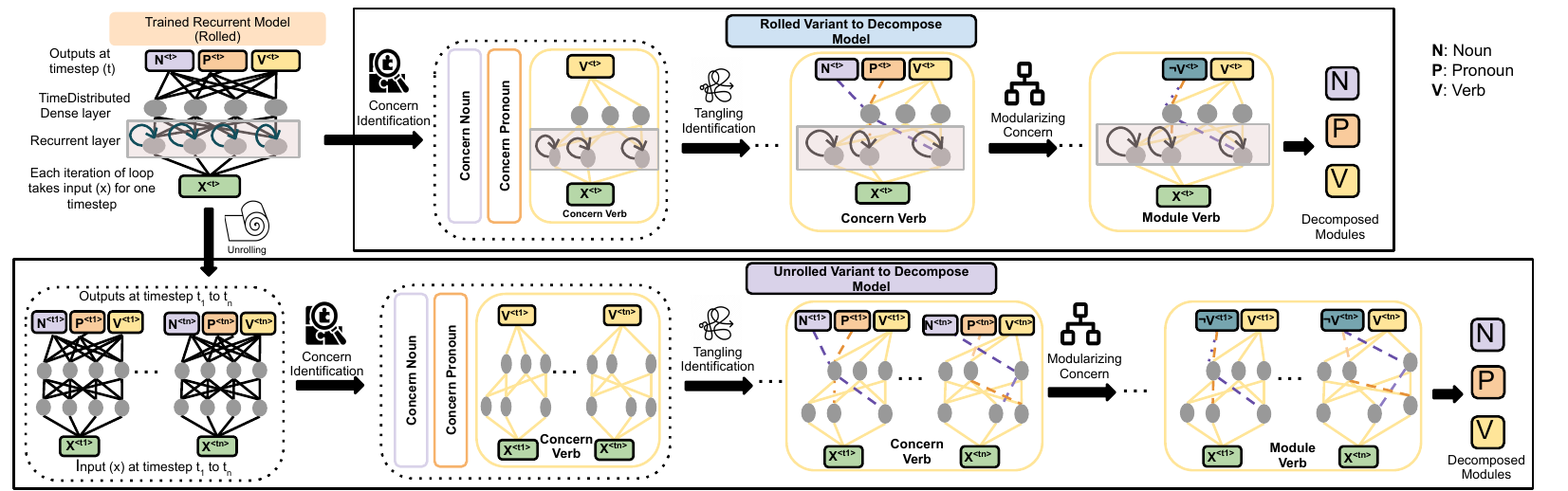}
	\vspace{-15pt}
	\caption{High-level overview of our approach.}
	\label{fig:overviewMany}
	%\vspace{14pt}
\end{figure*}
To introduce a new input (new language) to an existing RNN model, it must be trained from scratch. For instance, recently, the European Commission (EC) presented a new requirement to support the translation of languages that are not part of the European Union (EU)~\cite{multilingual, ukraine}.
It needs to be retrained to support the additional language. This process is expensive and time-consuming as developers must preprocess the data and train a model. \figref{fig:motivation} on the top right section shows how decomposition could help in this regard. 
Here, we have a monolithic model that translates an input in English into French, Italian, and German. Traditionally, a new model needs to be trained by adding new examples to the existing dataset to add the translation of English to the Ukrainian language.

In contrast, using our approach, one can take the existing trained model and decompose it into modules, in which each module is responsible for translating English to another non-English language. Then, find a different multi-lingual model trained to translate multiple languages, including English to Ukrainian, and decompose it. Next, take the module responsible for translating English to Ukrainian and compose it with the existing model's modules. Moreover, one can also train a small model that only supports English to Ukrainian translation. Then, compose it with the decomposed modules from the existing model. In both cases, training the large model can be avoided. The choice of approach depends on the available resources and module(s).

\textbf{Altering a part of a trained model.} 
Like any software system, neural network models can exhibit faulty behavior. For instance, in Estonian, ``ta on arst'' means ``She is a doctor'' in English. However, since Estonian is a gender-inclusive language, when we used the Google translator~\cite{google}, the output was ``He is a doctor'', which is incorrect and gender-biased. In such scenarios, the European Language Resource Coordination (ELRC) is set to actively enhances the multilingual translation services of EU languages~\cite{multilingual}. The most common approach to updating existing models is two-pronged: 1) introducing new examples to improve the faulty data, and 2) retraining the whole model. However, this approach can be resource-intensive because of the retraining.

In contrast, using our approach, one can handle the same problem in two different ways (\figref{fig:motivation} bottom right part). In the first approach, we decompose the existing trained model into modules. Then, we remove the Estonian to English translation module. Lastly, we use decomposition to select a replacement from a non-biased Estonian to English model. 
We retrain a small model using an improved dataset in the second approach. Then we compose it with the previously decomposed modules. 
In both cases, retraining the large model can be avoided. Thus, saving computational resources and hardware costs.
\section{Related works}
\label{sec: related}

In the SE community, a vast body of work in software decomposition~\cite{parnas1972criteria, liskov1974programming, parnas1976design, dijkstra1982role, cardelli1997program} exists. The notion of modularity exists in the ML community too~\cite{andreas2016neural, hu2017learning, hinton2000learning, sabour2017dynamic, ghazi2019recursive}, however, to address different issues than what this work aims to deliver. They eventually produce a monolithic model in the sense that it does not enable fine-grained reuse. 

Many studies reuse a DNN model to solve a software engineering task~\cite{gao2021automating, chai2022cross, lin2021traceability}. Transfer learning is a common technique to reuse the knowledge and structure of a trained model~\cite{ pratt1991direct, zhang2020adapnet}. However, retraining and modification are required when applying transfer learning to a different task. Moreover, replacing the model's logic cannot be achieved by transfer learning. 

Along this line, the closest work was introduced by \citeauthor{pan2022decomposing}~\cite{pan2020decomposing, pan2022decomposing}. They propose an approach to decompose an FCNN and CNN multi-class model into modules, which can be (re)used with other module(s) or be replaced by other modules to solve various problems. 
\citeauthor{sairam2018hsd}~\cite{sairam2018hsd} identified sections of a trained CNN model to reuse when the target problem requires a set of output classes, i.e., the subset of the original model.
In contrast, our decomposed modules can be reused and replaced with modules originating from the same and different datasets.

While these works introduce the notion of decomposition in various DL models, they cannot be directly applied to RNNs for the following reasons: (a) loops in the architectures, (b) logistic activations, and (c) different I/O modes. In contrast, our work addresses these novel technical challenges and proposes an approach to decompose RNN models into modules.

\section{Approach}
\label{sec:approach}

In this paper, we propose a decomposition technique for recurrent models, where a binary module is produced for each output label. 
Figure~\ref{fig:overviewMany} shows the overall approach. It starts with a trained model and produces decomposed modules. We propose two variants (\rolled and \unrolled) to that end. Each variant uses a different strategy to handle the loops in the RNN architecture. 
Like prior work~\cite{pan2020decomposing, pan2022decomposing}, broadly, the process of RNN decomposition has three steps -- Concern Identification (CI), Tangling Identification (TI), and Concern Modularization (CM). First, CI identifies the model parts contributing to a concern (an output class). After CI, a model can mostly recognize the target output class. Therefore, TI aims to add/update some parts responsible for negative output classes. Lastly, CM modularizes the concerns and creates a module that can recognize a single output class. Next, we describe technical challenges first, then each step in detail. 

\subsection{Challenge}
\label{subsec:challenge}

Previous studies have proposed a decomposition approach for FCNN and CNN~\cite{pan2020decomposing,pan2022decomposing}. However, RNN models significantly differ from other model types rendering prior approaches inapplicable to RNNs. 
This section discusses such differences that our approach addresses.

\begin{figure*}[!htp]
    \centering
	\includegraphics[width=0.8\linewidth]{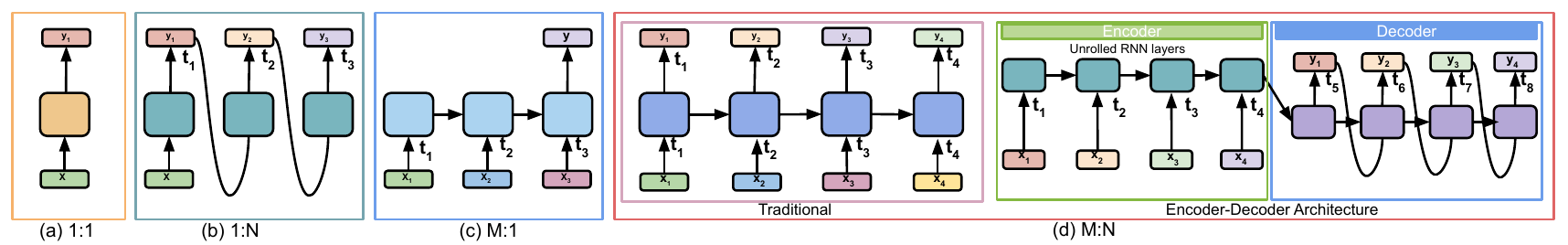}
% 	\vspace{-5pt}
	\caption{Different RNN input/output (I/O) architectures.}
	\label{fig:rnncell}
\end{figure*}

\textbf{Challenge 1: Weight Sharing Across Time.}
Traditional neural networks (i.e., CNNs and FCNNs) do not have loops in their architecture -- data directly flows from the input to the output layer. However, for RNNs, loops are leveraged to process sequential data. 
By doing so, the weights and biases are shared across time. Therefore, the decomposition algorithm must be aware of the time dimension in the network.

\textbf{Challenge 2: Additional Learned Parameters.}
Unlike traditional neural networks, every \rnn cell (node) receives two sets of learnable weights, i.e., weights associated with input at a timestep, $W$, and internal recurrent state or memory weights, $U$. 
The decomposition algorithm must decompose these additional learned parameters associated with memory.

\textbf{Challenge 3: Gated layer architecture.}
Improved RNN variants, such as LSTM and GRU, perform gate operations, which enable both long and short-term memory, unlike vanilla RNN, which only involves short-term memory. For example, LSTM involves three gates in its internal architecture: input, forget, and output gate. Internal operations of an LSTM cell is shown in $lstm\_op$ method in algorithm~\ref{algo:commonalgo}. Each of these gates incorporates learnable (weights and biases) parameters of its own. Hence, the decomposition technique must consider these gates, i.e., identify relevant nodes across all gates.

\textbf{Challenge 4: New activation function.}
Previous works on DNN decomposition only support \relu activation function~\cite{pan2020decomposing, pan2022decomposing}, which are common in FCNN and CNN. However, in RNNs, logistic activation functions, such as \textit{Tanh, Sigmoid}, are the most common~\cite{cho2014properties,hochreiter1997long,lipton2015critical}. Moreover, popular DL libraries such as \textit{Keras, TensorFlow, and PyTorch} use \textit{Tanh} as the default activation function for RNNs~\cite{kerasrnn,tfrnn,torchrnn}. Therefore, to apply decomposition, we must support these functions.

\textbf{Challenge 5: Multiple I/O Architecture.}
FCNN and CNN models process a single output from one given input (\textit{1:1}). As a result, prior works~\cite{pan2020decomposing, pan2022decomposing} were proposed for networks with a \textit{1:1} architecture. However, there are 5 different input-output architectures in RNNs, as shown in Figure~\ref{fig:rnncell}, which also need to be supported. \textbf{1:1} models receive a single input and produce one output. \textbf{1:N} models take a single input and produce sequential outputs. 
\textbf{M:1} models receive sequential inputs and produce one output. 
\textbf{M:N} models receive a sequence of inputs and produce many outputs. Figure~\ref{fig:rnncell} shows the two sub-types for \textit{M:N} models. On the right, the architecture takes inputs and produces outputs at different timesteps (e.g., encoder-decoder models). The left one produces the output at the same timestep as it receives its inputs.

\subsection{Concern Identification}
\label{subsec:ci}

Concern identification (CI) aims to identify parts of the network (nodes and edges) responsible for classifying an output label, $OL$, referred to as a concern. The definition of concern is aligned with the traditional SE and prior works on decomposing DNN into modules~\cite{pan2020decomposing, pan2022decomposing}. The output label for which the concerned nodes and edges are identified is the dominant output class in the concern, while other classes are non-dominant. 
For instance, in Figure~\ref{fig:overviewMany}, for the \textit{concern verb}, verb (\textit{V}) is the dominant class, and pronoun (\textit{P}) and noun (\textit{N}), are the non-dominant classes. In that example, we show how a model, taking many inputs (words) and tagging each word with a part-of-speech (POS), can be decomposed into modules for each POS. 

On a high level, CI aims to identify nodes and edges relevant to a particular concern or output class. In a ReLU-based network, relevant nodes can be identified by observing activation values for a sample of that class, and nodes that always remain active can be treated as relevant. However, because of logistic activation used in RNNs, where the notion of "active" or "inactive" is not as distinct as in ReLU, we identify relevant nodes by comparing the central activation level of a node in both positive and negative samples. Besides that, CI for RNN must also be aware of the time loop, multi-output, and gated architecture of RNNs. Our approach unrolls the time loop in RNN and performs CI (i.e., identifying relevant nodes) in both a time-sensitive (unrolled) and insensitive manner (rolled). Moreover, improved RNN variants (i.e., LSTM) involve gates in its architecture, which we handle by identifying relevant nodes across all gates. Finally, a different CI approach is needed for multi-output models from a single one, as one input example can be labeled with multiple classes (concerns). While CI for single output can monitor a single input sample for one particular class, we must enable simultaneous detection of multiple concerns for multi-output models. Next, we discuss our approaches to handling these challenges while decomposing an RNN model in detail.

% briefly discuss how we handle challenges outlined in \S~\ref{subsec:challenge} to decompose an RNN model.

\subsubsection{Identifying Concerns Across Time}
First, in the RNN models, the weights are shared across timesteps (in RNN notations, each point in time is called a \textit{timestep}). As a result, different nodes may activate at distinct timesteps. Therefore, in our approach, we propose two CI variants. In particular, the \textit{rolled}, which does not take the timesteps of a node into account, and the \textit{unrolled}, which is timestep-sensitive.  

\begin{algorithm}[!htp]
	\caption{Decomposition:\{One, Many\}-to-One Models.}
	\scriptsize
	\label{algo:one}
	%\footnotesize
	\begin{algorithmic}[1]
	
    	\Procedure {One($model, activation, rolled, input\_timestep, X,Y$)}{}
            \State{$modules=[\:]$}\label{algoone:1}
            \For {$every \: output \: label, ol$}\label{algoone:2}
                \State $p\_in~=~sample(X,Y, only~=~ol, size~=~M)$\label{algoone:3}
                \State  $n\_in~=~sample(X,Y, not~=~ol, size=M/|class|-1)$\label{algoone:4}
                \State {$concern\_o=initConcern(model,timestep,rolled)$}\label{algoone:init}

                %\Comment {Do observe positive and negative examples and calculate central activation tendency, CT, for each neuron}\label{algoone:5}
                \State {$h\_val\_pos=monitor(p\_in, model, rolled)$}\label{algoone:6}
                \State {$h\_val\_neg=monitor(n\_in, model, rolled)$}\label{algoone:7}
                
                \If {$rolled$}\label{algoone:8}
                    \State {$flat\_p=flatten\_obs(model, h\_val\_pos, timestep)$}\label{algoone:9}
                    %\Comment{treat each timestep as separate observation}
                    \State{$flat\_n=flatten\_obs(model,h\_val\_neg,timestep)$}\label{algoone:10}
                  \State {$update\_concern(concern\_o, flat\_p, flat\_n, activation)$}\label{algoone:11}
                \Else
                \For {$ts~\in~input\_timestep$}\label{algoone:12}
                    \State{$p\_ts=obs\_at(model,h\_val\_pos, ts)$}\label{algoone:13} %\Comment{get hidden values for all observations at ts}
                    \State $n\_ts=obs\_at(model,h\_val\_neg, ts)$\label{algoone:14}
                    \State $con\_ts =initConcern(model,timestep,True)$
                    \State $update\_concern(con\_ts, p\_ts, n\_ts, activation)$
                    \State $merge(concern\_o, con\_ts, ts)$\label{algoone:15}
                \EndFor
                \EndIf
                \State{$cur\_module=channel(concern\_o)$}\label{algoone:16}
              \State{$append(modules, cur\_module)$}\label{algoone:17}
            \EndFor
            \Return $modules$
        \EndProcedure

	\end{algorithmic}    
\end{algorithm}
%\vspace{-10pt}

\textbf{Rolled:} In this variant, we identify the concern and remove the unrelated nodes and edges. Instead of keeping multiple copies of the updated edges and nodes (one for each timestep), we keep a single copy of the modified edges and nodes. All timesteps share the same weights obtained after removing edges and nodes.

\textbf{Unrolled:} In traditional SE, to understand the impact of each loop iteration, prior works have proposed approaches~\cite{dongarra1979unrolling, weiss1987study, davidson1995improving} to unroll the loops. Inspired by such works, the unrolled variant follows a two-step concern identification process. In particular, it first unrolls the RNN model loop while transforming it into an equivalent sequence of operations. Then, it identifies the nodes and edges responsible for each concern at each timestep. 

\begin{algorithm}[!htp]
	\caption{Decomposition: \{One, Many\}-to-Many Models.}
	\scriptsize
	\label{algo:many}
	\begin{algorithmic}[1]
    
    	    \Procedure {Many($model, activation, rolled,output\_timestep,X,Y$)}{}
            \State{$modules=[\:]$}\label{algomany:1}
            \For {$every \: output \: label, ol$}\label{algomany:2}
                \State {$concern\_o=initConcern(model,timestep,rolled)$}
                \State {$flat\_p=[]$}\label{algomany:4}
                \State {$flat\_n=[]$}\label{algomany:5}
                
                \For {$every \: output\_timestep, ts$}\label{algomany:6}
                \algorithmiccomment{monitor one step at a time}
                    \State {$p\_in=sample(X,Y, ts, only=ol, size=M)$}\label{algomany:7}
                    \State {$n\_in=sample(X,Y, ts, not=ol, size=M/|class|-1)$}\label{algomany:8}
                %\Comment{ Do observe positive and negative examples and calculate  central activation tendency, CT, for each neuron}
                    \State {$h\_val\_pos = monitor(p\_in, model, rolled)$}\label{algomany:9}
                    \State {$h\_val\_neg = monitor(n\_in, model, rolled)$}\label{algomany:10}
                    \State {$p\_ts = obs\_at(h\_val\_pos, ts)$} \label{algomany:11}
                    %\Comment get hidden values for all observations at ts, ignore effect of these samples on other timesteps
                    \State {$n\_ts = obs\_at(h\_val\_neg, ts)$} \label{algomany:12}
                    
                    \If {$rolled$}\label{algomany:13}
                        \State{$concat(flat\_p, h\_val\_pos\_ts, axis=0)$}\label{algomany:14}
                        \State {$concat(flat\_n, h\_val\_neg\_ts, axis=0)$}\label{algomany:15}
                    \Else 
                         \State {$con\_ts =initConcern(model,timestep,True)$}\label{algomany:inittemp}
                         \State {$update\_concern(con\_ts, p\_ts, n\_ts, activation)$}
                        %  \State  merge(concern\_o, con\_ts, ts)\label{algomany:15}
                         \State {$concern\_o[ts].W, U, B=con\_ts.W, U, B$}\label{algomany:16}\algorithmiccomment{Merge}

                    \EndIf
                \EndFor
                \If {$rolled$}\label{algomany:17}
                    \State {$update\_concern(concern\_o, flat\_p, flat\_n, activation)$} \label{algomany:18}
                \EndIf
                \State {$cur\_module=channel(concern\_o)$} \label{algomany:19}
                \State {$append(modules, cur\_module)$}\label{algomany:20}
            \EndFor
            \Return modules

        \EndProcedure

	\end{algorithmic}    
\end{algorithm}
%\vspace{-10pt}

\subsubsection{Support logistic activation}
\label{subsec:suplog}
Unlike \relu, logistic activations squash the given input within a certain range~\cite{salehinejad2017recent}. For example, Tanh squashes input in range $[-1,+1]$, and Sigmoid in range $[0,1]$~\cite{salehinejad2017recent}. In that regard, Lipton~\cite{lipton2015critical} \etal reported that the hidden layers of a model using logistic activation are rarely sparse, unlike \relu-based layers. This is because a node value is rarely exactly zero for logistic activation-based layers, unlike \relu-based ones~\cite{lipton2015critical}. As a result, prior \relu-based decomposition techniques, which leverages the notion of ``on'' or ``off'' (``off'', if node value is 0, otherwise, ``on'') to identify concerned nodes, do not apply to the logistic activation functions.

\begin{algorithm}[!htp]
	\caption{Helper algorithms for algorithm~\ref{algo:one} and \ref{algo:many}}
	\scriptsize
	\label{algo:commonalgo}
	%\footnotesize
	\begin{algorithmic}[1]
		\Procedure{initConcern ($model$, $timestep$, $rolled$)}{}
		
		\State $con: object$\algorithmiccomment{Initialize new model object}

		\For{$each~layer, l  \in model$}\label{algoinit:3}
		  \If {$layer.generic\_type==``Recurrent"~and~not~rolled$}\label{algoinit:4}

		      %  \If{$not~rolled$}\label{algoinit:5}
		          % \State $con[l].W,con[l].U,con[l].B=[\:]$\label{algoinit:6}
		            \For{$each~ts  \in timestep$}\label{algoinit:8}
						\State {$con[l].W[ts],U[ts],B[ts]~=~l.get\_weights()$}\label{algoinit:9}
					\EndFor

			\Else\label{algoinit:4}

    					\State $con[l].W,U,B=l.get\_weights()$\label{algoinit:12}\algorithmiccomment{U=N/A if dense layer}
		    \EndIf
		        
		      %  \Else
		      %  	    \State $con[l].W,U=l.get\_weights()$\label{algoinit:15}

		\EndFor
		
		 \Return $concern$
		
		\EndProcedure
		
		\Procedure {monitor($samples$, $model$)}{}
    
            \State $hidden\_val=[\:]$\label{algoma:1}
            \For {$x \in samples$}\label{algoma:2}
                \For {$each \: layer \in model$}\label{algoma:3}
                    \If {$layer.generic\_type==``Recurrent"$}\label{algoma:4}
                        \For {$ts \in timestep$}\label{algoma:lstmloop}
                            \If {$layer.type==``LSTM"$}
                                \State {$h_t,c_t=lstm\_op(layer, x, h_{t-1}, c_{t-1})$}\label{algoma:lstmop}
                            \EndIf
                            \State ...
                            % \If {$layer.type==``GRU"$}
                            %     \State {$h_t=gru\_op(layer, x, h_{t-1})$}\label{algoma:gruop}
                            % \EndIf
                            %  \If {$layer.type==``RNN"$}
                            %     \State {$h_t=vanilla\_rnn\_op(layer, x, h_{t-1})$}\label{algoma:rnnop}
                            % \EndIf
                            \State {$append (hidden\_val, ts, layer, h_t)$}\label{algoma:lstmrecord}
                        \EndFor
                    \EndIf
                    
                    \If {$layer.type==``Dense"$}\label{algoma:8}
                        \State {$h_t = x.dot(layer.W)+layer.B$}\label{algoma:9}
                        \State {$append (hidden\_val, layer, h_t)$}
                    \EndIf
                \EndFor
            \EndFor
            \Return $hidden\_val$
        
        \EndProcedure
        
        \Procedure {flatten\_obs($model, hidden\_val, timestep$)}{}
           \State {$flattened=[\:]$}\label{algoft:1}

           \For {$s \in timestep$}\label{algoft:2}
                \For {$each~layer \in model$}\label{algoft:3}
                    \State {$append(flattened,layer, hidden\_val[ts][layer])$}\label{algoft:4}
                \EndFor
            \EndFor
		    \Return $flattened$
        \EndProcedure
        
         \Procedure {obs\_at($model, hidden\_val, ts$)}{}
           \State {$values=[\:]$}\label{algohva:1}

            \For {$each~layer \in model$}\label{algohva:3}
                \State {$append(values,layer, hidden\_val[ts][layer])$}\label{algohva:4}
            \EndFor
		    \Return $values$
        \EndProcedure
        
         \Procedure {update\_concern$(concern, h\_pos, h\_neg, activation)$}{}
            \If {$activation==``logistic"$}\label{algooc: 1}
                \State $ct\_pos=central\_tendency(h\_pos, model)$\label{algooc:2}
                \State $ct\_neg=central\_tendency(h\_neg, model)$\label{algooc:3}
                \State $CI\_logistic(ct\_pos, ct\_neg, concern)$ \algorithmiccomment{identify concern}\label{algooc:4}
            \EndIf
            \If{$activation==``relu"$}\label{algooc:5}
                \State $rate\_pos=active\_rate(h\_pos, model)$\label{algooc:6}
                \State $rate\_neg=active\_rate(h\_neg, model)$\label{algooc:7}
                \State $CI\_relu(rate\_pos, concern)$ \algorithmiccomment{identify dominant nodes}\label{algooc:8}
                \State $TI\_relu(rate\_neg, concern)$ \algorithmiccomment{identify tangling nodes}\label{algooc:9}
            \EndIf
		    \Return $concern$\algorithmiccomment{Return updated concern}
        \EndProcedure

        % \Procedure {merge($concern\_o, concern\_ts, ts$)}{}
        %   \State {$concern\_o[ts].W, U, B=concern\_ts.W, U, B$}\label{algomer:1}
        %   \State {$concern\_o[ts].U=concern\_ts.U$}\label{algomer:2}
        %   \State {$concern\_o[ts].B=concern\_ts.B$}\label{algomer:3}

        % \EndProcedure
		
		\Procedure{lstm\_op ($layer$, $x_{t}$,  $h_{t-1}$,$c_{t-1}$)}{}\algorithmiccomment{Apply LSTM Op.}
        
            \State {$s_t=x_t.dot(layer.W)+h_{t-1}.dot(layer.U)+layer.B$}
            \State $n=layer.num\_hidden\_neuron$
            \State $i,f,o=\sigma(s_t[:,:n]),\sigma(s_t[:,n:2n]),\sigma(s_t[:,3n:])$\label{algolstm:3}\algorithmiccomment{Gates.}
            % \State $f=\sigma(s_t[:,n:2n])$\label{algolstm:4}\algorithmiccomment{Forget gate.}
            % \State {$\_c=tanh(s_t[:,2n:3n])$}\label{algolstm:5}
            % \State {$o=\sigma(s_t[:,3n:])$}\label{algolstm:6}\algorithmiccomment{Output gate.}
            \State $c_{t}=i*tanh(s_t[:,2n:3n])+f*c_{t-1}$\label{algolstm:7}\algorithmiccomment{Cell state.}
            \State $h_{t}=o*tanh(c_{t})$\label{algolstm:8}\algorithmiccomment{Hidden state.}
    		\State \Return $h_{t},c_{t}$\label{algolstm:9}

		\EndProcedure
		
% 		\Procedure{gru\_op ($layer$, $x_{t}$,  $h_{t-1}$)}{}\algorithmiccomment{Apply GRU Op.}
        
%             \State {$w,u,b=layer.W,layer.U,layer.B$}
%             \State $n=layer.num\_hidden\_neuron$
%             \State $z=\sigma(x_t.dot(w[:,:n])+h_{t-1}.dot(u[:,:n])+b[:n])$\label{algogru:3}%\algorithmiccomment{Update gate.}
%             \State $r=\sigma(x_t.dot(w[:,n:2n])+h_{t-1}.dot(u[:,n:2n])+b[:2n])$\label{algogru:4}%\algorithmiccomment{Reset gate.}
%             \State $h=x_t.dot(w[:,2n:])+r*h_{t-1}.dot(u[:,2n:])+b[2n:]$\label{algogru:5}

%             \State $h_{t}=z*h_{t-1}+(1-z)*h$\label{algogru:6}\algorithmiccomment{Hidden state.}
%     		\State \Return $h_{t}$\label{algogru:7}

% 		\EndProcedure
		
% 		\Procedure{vanilla\_rnn\_op ($layer$, $x_{t}$,  $h_{t-1}$)}{}%\algorithmiccomment{Apply RNN Op.}
            
%             \State $h_t=f(x_t.dot(layer.W)+h_{t-1}.dot(layer.U)+layer.B)$\label{algornn:1}%\algorithmiccomment{obtain next RNN state.
            
% %             \If{$layer.activation==``relu"$}\label{algo3p1:12}
% % 			    \State $h_t=relu(h_t)$
% % 			\EndIf
% % 			 \If{$layer.activation==``tanh"$}\label{algo3p1:12}
% % 			    \State $h_t=tanh(h_t)$
% % 			\EndIf
									
%     		\State \Return $h_{t}$\label{algolstm:9}

% 		\EndProcedure

	\end{algorithmic}    
\end{algorithm}

To address this issue, our intuition is that a logistic activation value can be regarded as the excitation level of a node. A higher activation value indicates to what extent it lets information through that node. For instance, an activation value 1 for \textit{Sigmoid} allows maximum information through that node, while 0.0 is the lowest. Leveraging this insight, we compare the central activation tendency of a neuron in positive and negative examples and identify concerns (Algorithm - ~\ref{algo:citanh}). To that end, first, we sample a set of positive (inputs with target output label) and negative examples (inputs with other output labels). Then, we monitor the activation of nodes for both sets while comparing their central activation tendency. The method $central\_tendency$ measures the central activation tendency of a neuron. First, it retrieves the distribution of activation values for the observed examples (Line~\ref{algoct:3} of Algorithm~\ref{algo:citanh}) for a neuron, $node$. Note that the absolute value of observed activations is taken (Line~\ref{algoct:abs} of Algorithm~\ref{algo:citanh}) as \textit{Tanh}-activated values could lie in the range [-1,+1] and -1 allows as much information as +1, but in the opposite direction. Then, it computes the mean as a measure of central activation tendency for that neuron after discarding outlier observations (Line~\ref{algoct:9} of Algorithm~\ref{algo:citanh}). 
In method $CI\_logistic$, a node is considered more relevant to the dominant output label if its central tendency is higher in positive examples than in negative ones (Line ~\ref{algocitanhp1:7} of Algorithm~\ref{algo:citanh}). 
Because it tends to allow more information to pass through for the dominant class than for other classes, we keep this node and edges going in/out of it for the dominant class. The method stops removal if the graph becomes too sparse. To that end, taking the 10\% removal threshold used by Pan \etal~\cite{pan2020decomposing} as a starting point, we evaluate the decomposition quality at a 5\% interval (i.e., 10\%, 15\%, 20\%, etc.). For RNN models, we found that beyond 20\%, the graph starts to become too sparse, affecting decomposition adversely, and we selected this as the threshold in our experiments.

\subsubsection{Support gated layer}
To support the decomposition of gated layer architecture, we propose two approaches: i) decompose each gate based on its own activation, and ii) decompose all gates based on LSTM/GRU final hidden state. Note that hidden states or values refer to the output of the model's intermediate or hidden layers (between input and output layers). In the first approach, we identify dominant nodes in each gate based on their own activation values, while, in the latter, we remove a node from all gates if its corresponding hidden state value is found to be less significant in positive samples. The hidden state at a particular timestep dictates the output at that step. As a result, a node's hidden value indicates its relevance to the output produced at that step. Therefore, in the second approach, we compare a node's central hidden state tendency for positive and negative examples. We remove it from each gate if found to be more relevant to non-dominant classes or negative examples  (Line~\ref{algocitanhp1:7}-\ref{algocitanhp1:13} in Algorithm~\ref{algo:citanh}).

\begin{algorithm}[!htp]
	\caption{Concern identification for logistic activation.}
	\scriptsize
	\label{algo:citanh}
	%\footnotesize
	\begin{algorithmic}[1]
	
	    \Procedure{CI_logistic $(ct\_pos$, $ct\_neg$, $concern$, $thres)$}{}
		    
		    \State $d=\{\}$\label{algocitanhp1:1}
		    
		    \For{$each~ layer, l \in concern$}\label{algocitanhp1:2}
    			\For{$each~ node, n \in l$}\label{algocitanhp1:3}
    			    \State $d[l][n]=ct\_pos[l][n]-ct\_neg[l][n]$\label{algocitanhp1:4}
    			\EndFor
			\EndFor
			
			 \State $d=sort(d, order=``ascending")$\label{algocitanhp1:5}
			 
			 \For{$each~layer,node~\in~d.keys()$}\label{algocitanhp1:6}
			    	\If{$d[layer][node]<0.0$}\label{algocitanhp1:7}
						\State $removeNode(concern, layer,node)$\label{algocitanhp1:8}
						\If{$layer.type=``LSTM" or layer.type=``GRU"$}\label{algocitanhp1:9}
						    \State $hunit=layer.num\_hidden\_neurons$\label{algocitanhp1:10}
						    \State $removeNode(concern, layer,hunit+node)$\label{algocitanhp1:11}
						    \State $removeNode(concern, layer,2*hunit+node)$\label{algocitanhp1:12}
						    \If{$layer.type=``LSTM"$}\label{algocitanhp1:onlylstm}
						    \State $removeNode(concern, layer,3*hunit+node)$\label{algocitanhp1:13}
						    \EndIf
					    \EndIf
					\EndIf
					
					\If{$removedPercent>=thres$}\label{algocitanhp1:14}
						\State $stop$
					\EndIf
			\EndFor
		
		    \Return $concern$\algorithmiccomment{Return updated concern}
		\EndProcedure
	    
% 	    \Procedure{CI_logistic $(ct\_pos$, $ct\_neg$, $concern$, $removal\_threshold)$}{}
		    
% 		    \State d=\{\:\}\label{algocitanhp1:1}
		    
% 		    \For{each \: layer, l $\in$ concern}\label{algocitanhp1:2}
%     			\For{each \: neuron, n $\in$ l}\label{algocitanhp1:3}
%     			    \State d[l][n]=ct\_pos[l][n]-ct\_neg[l][n]\label{algocitanhp1:4}
%     			\EndFor
% 			\EndFor
			
% 			 \State d=sort(d, order=``ascending")\label{algocitanhp1:5}
			 
% 			 \For{each \: layer,neuron  $\in$ d.keys()}\label{algocitanhp1:6}
% 			    	\If{d[layer][neuron]$<$0.0}\label{algocitanhp1:7}
% 						\State removeNeuron(concern, layer,neuron)\label{algocitanhp1:8}
% 						\If{layer.type=LSTM or layer.type=GRU}\label{algocitanhp1:9}
% 						    \State hunit=layer.num\_hidden\_neurons\label{algocitanhp1:10}
% 						    \State removeNeuron(concern, layer,hunit+neuron)\label{algocitanhp1:11}
% 						    \State removeNeuron(concern, layer,2*hunit+neuron)\label{algocitanhp1:12}
% 						    \If{layer.type=LSTM}\label{algocitanhp1:onlylstm}
% 						    \State removeNeuron(concern, layer,3*hunit+neuron)\label{algocitanhp1:13}
% 						    \EndIf
% 					    \EndIf
% 					\EndIf
					
% 					\If{removedPercent$>=$removal\_threshold}\label{algocitanhp1:14}
% 						\State stop
% 					\EndIf
% 			\EndFor
		
% 		    \Return concern\algorithmiccomment{Return updated concern}
% 		\EndProcedure

\Procedure {central_tendency($hidden\_val, concern, activation$)}{}
    
    \State $tendency=\{\}$\label{algoct:1}
    \For {$each~layer,node \in concern$}\label{algoct:2}
        \State $all\_obs=[]$
        \For {$each~observation, o \in hidden\_val$}\label{algoct:3}
            % \If{activation=Tanh}\label{algoct:4}
			    \State $all\_obs.append(abs(hidden\_val[layer][node][o]))$\label{algoct:abs}
% 			\EndIf
% 			\If{activation=Sigmoid}\label{algoct:6}
% 			    \State all\_obs.append(hidden\_val[layer][n][o])\label{algoct:7}
% 			\EndIf
            \State 
        \EndFor

        % \State {$d~=~math.abs(all\_obs - math.median(all\_obs))$}\label{algoct:5}
        % \State{$mdev~=~np.median(d)$}\label{algoct:6}
        % \State{$s = d / mdev~if~mdev~else~0$}\label{algoct:7}
        % \State{$d = all\_obs[s < m]$}\label{algoct:8}
        \State{$d = remove\_outliers(all\_obs)$}\label{algoct:8}
        \State $tendency[layer][node]=mean(d)$\label{algoct:9}
    \EndFor
    \Return $tendency$
\EndProcedure		

% \Procedure {central_tendency($hidden\_val, concern, activation$)}{}
    
%     \State tendency=\{\:\}\label{algoct:1}
%     \For {each neuron n $\in$ concern}\label{algoct:2}
%         \State all\_obs=[\:]
%         \For {each observation o $\in$ hidden\_val}\label{algoct:3}
%             % \If{activation=Tanh}\label{algoct:4}
% 			    \State all\_obs.append(abs(hidden\_val[layer][n][o]))\label{algoct:5}
% % 			\EndIf
% % 			\If{activation=Sigmoid}\label{algoct:6}
% % 			    \State all\_obs.append(hidden\_val[layer][n][o])\label{algoct:7}
% % 			\EndIf
%             \State 
%         \EndFor

%         \State {d = math.abs(all\_obs - math.median(all\_obs))}\label{algoct:5}
%         \State{mdev = np.median(d)}\label{algoct:6}
%         \State{s = d / mdev \: if \: mdev \: else \: 0}\label{algoct:7}
%         \State{d = all\_obs[s $<$ m]}\label{algoct:8}
%         \State tendency[n]=math.mean(d)\label{algoct:9}
%     \EndFor
%     \Return tendency
% \EndProcedure

	\end{algorithmic} 
\end{algorithm}

\subsubsection{Support Multiple I/O Architectures}
\label{subsec:multiarch}
To support the CI in models with multiple input and output classes (i.e.,~\textit{1:N}, and \textit{M:N}), we propose two approaches for two different output modes, one and many. Algorithm~\ref{algo:one} and ~\ref{algo:many} show the details as described next.

\textbf{\{One, Many\}-to-One.} Like any traditional network, \rnn can take a single input and produce a single output. However, \rnn can also take many inputs (i.e., sequential data). Algorithm~\ref{algo:one} shows our approach for concern identification in the presence of an input loop and models that produce a single output. Algorithm~\ref{algo:one} receives a trained model, activation type, modularization mode (i.e., \rolled or \unrolled), \tss (1 for \one, $>1$ for \many at input end), and input examples. Next, it iterates through every output label ($OL$) to create a module for each output label. 
To identify the concern for $OL$, first, our approach selects the examples labeled as the dominant output class in the training dataset (Line~\ref{algoone:3} of Algorithm~\ref{algo:one}), which we call a positive sample. Also, it selects negative samples proportionally from each negative class (Line~\ref{algoone:4} of Algorithm~\ref{algo:one}).

Then, in line~\ref{algoone:init} of algorithm~\ref{algo:one}, \textit{initConcern} method initializes the modular weights. In particular, it initializes the modular weights \textit{W}, \textit{U}, and \textit{B} with the trained model ones. These nodes and edges from the weights are removed and updated as the algorithm proceeds to identify a concern. In the \textit{unrolled} mode, for \textit{recurrent} layers, \textit{initConcern} creates \ts copies of \textit{W}, \textit{U} and \textit{B} to allow individual timestep-specific weight pruning (Line~\ref{algoinit:8}-\ref{algoinit:9}) in Algorithm~\ref{algo:commonalgo}).

In the next step, each neuron activation is monitored for both positive and negative samples (Line~\ref{algoone:6}-\ref{algoone:7} of Algorithm~\ref{algo:commonalgo}) by feeding them to the trained model. $monitor$ method of the Algorithm~\ref{algo:commonalgo} shows how the nodes are observed. In particular, each example is propagated through the trained model while recording the node hidden values for each example across all timesteps. The method handles different types of trainable layers found in a recurrent model. For recurrent layers, the method handles the presence of a loop by implementing a feedback loop (Line~\ref{algoma:lstmloop} of Algorithm~\ref{algo:commonalgo}). For example, for LSTM layers, LSTM cell, $lstm\_op$, is repeatedly fed with an input, $x_t$, at a particular \ts, \textit{t}, previous hidden and cell state (Line~\ref{algoma:lstmop} of Algorithm~\ref{algo:commonalgo}). The cell performs a stateful input transformation, $x_t$, by using the contextual information from the previous cell and recording hidden values at each timestep (Line~\ref{algoma:lstmrecord} of Algorithm~\ref{algo:commonalgo}). 

\begin{algorithm}[!htp]
	\caption{Concern identification for \relu.}
	\scriptsize
	\label{algo:cirelu}
	%\footnotesize
	\begin{algorithmic}[1]
	
	    \Procedure {active\_rate($hidden\_val$, $concern$)}{}
    
            \State $rate=\{\}$\label{algoar:1}
            \For {$each~node,n~\in~concern$}\label{algoar:2}
                \State $activeCounter=0$
                \For {$each~observation,o~\in~hidden\_val$}\label{algoar:3}
                    \If {$isNodeActive(hidden\_val[layer][n][o])~=~True$}\label{algoar:4}
                    \State $activeCounter+=1$ 
                    \EndIf
                \EndFor
        
                \State {$rate[n]=(activeCounter/len(hidden\_val))*100.0$}\label{algoar:5}
            \EndFor
            \Return $rate$
        \EndProcedure
        
        \Procedure{CI\_relu($active\_rate,concern, thres)$}{}
		  
            \For{$each~node~n~\in~concern$}\label{algocir:1}
                \If {$active\_rate[n]$==$0.0$}\label{algocir:2}
					    \State $removeNode(concern, layer,n)$\label{algocir:3}
                \EndIf
                \If {$removedPercent>=thres$}\label{algocir:4}
                    \State $stop$
                \EndIf
            \EndFor
		    \Return $concern$\algorithmiccomment{Return updated concern}

        \EndProcedure

        \Procedure{TI\_relu($active\_rate,concern)$}{}
        		  
            \For{$each~node,n~\in~concern$}\label{algotir:1}
                \If {$active\_rate[n]>0.0$}\label{algotir:2}
					    \State $restoreNode(concern, layer,n)$\label{algotir:3}
                \EndIf
            \EndFor
		    \Return $concern$\algorithmiccomment{Return updated concern}
        
        \EndProcedure

    \end{algorithmic} 
\end{algorithm}

\textit{Rolled}-variant of the algorithm identifies concern for the dominant class in a timestep-insensitive manner. Instead of identifying concerns in each timestep separately, all timesteps share the same identified nodes and edges for an output class. To that end, method $flatten\_obs$ is called to flatten observations/hidden values from all timesteps (Line~\ref{algoone:9} of Algorithm~\ref{algo:one}). This method essentially treats observed activation values of nodes at each timestep as a distinct observation of its own. For example, consider a model with timestep, 10, and 100 input samples to observe. In this case, each neuron will have $100*10$ observations after invoking $monitor$. In particular, it will observe a neuron, X, 100 times in each timestep. However, \textit{rolled}-variant will treat hidden values from different timesteps for a neuron as a separate observation as if there were $100*10$ input samples. Hence, in this mode, each neuron will have 1000 observations. 

Then, the concern is updated based on the 1000 observations in the $update\_concern$ method (Line~\ref{algoone:11} of Algorithm~\ref{algo:one}). For logistic activations, in $update\_concern$, it first computes the central activation tendency for these 1000 observations (Line~\ref{algooc:2}-\ref{algooc:3} of Algorithm~\ref{algo:commonalgo}). Then, it identifies relevant nodes and edges for the current concern, $OL$. Similarly, for \relu, it computes the active percent of a node given these observations and identifies relevant nodes accordingly (Line ~\ref{algooc:6}-\ref{algooc:9} of Algorithm~\ref{algo:commonalgo}). In particular, we keep nodes that are observed to be always \textit{active} (for Relu-based models) or comparatively more intensely activated (for logistic-based models); otherwise, removed as shown in Algorithm \ref{algo:citanh} and \ref{algo:cirelu}.

However, in \textit{unrolled} mode, the algorithm is timestep-sensitive as it goes to identify dominant nodes at each timestep separately (Line ~\ref{algoone:12}-\ref{algoone:15} of Algorithm~\ref{algo:one}). It iterates through each timestep and retrieves observations at that timestep (Line ~\ref{algoone:13},\ref{algoone:14} of Algorithm~\ref{algo:one}). Then, it creates a temporary object, $con\_ts$, to represent concern at this particular timestep. Finally, it merges the identified relevant nodes at this timestep to the concern under analysis, $conern\_o$, by adding them at that timestep (line~\ref{algoone:15} of Algorithm~\ref{algo:one}).

\textbf{\{One, Many\}-to-Many.} Unlike \textit{\{One, Many\}-to-One}, an input example can be associated with multiple output classes in \textit{\{One, Many\}-to-Many} models. For instance, the POS-tagging example shown in Figure~\ref{fig:overviewMany} has many outputs. In this example, every individual word in a given sentence is associated with a POS tag in the output. This kind of many-output problem poses unique challenges to the decomposition technique proposed for DNNs in the past~\cite{pan2020decomposing,pan2022decomposing}. The one-output-based technique can uniquely monitor a single input sample for one particular output label, as shown in Algorithm~\ref{algo:one}. However, it is not possible for many-output models as multiple concerns may be present simultaneously in a single input. Therefore, to decompose such models, our insight is to monitor each output timestep at a time, as shown in Algorithm~\ref{algo:many}.

The Algorithm~\ref{algo:many} starts by receiving the same parameters as Algorithm~\ref{algo:one}. Then, to build one module for each output class, it identifies dominant nodes in each timestep separately (Line~\ref{algomany:6} of Algorithm~\ref{algo:many}). For each output timestep, it similarly samples positive and negative examples to Algorithm~\ref{algo:one}. However, the sampling procedure is timestep-sensitive (Line ~\ref{algomany:7} of Algorithm~\ref{algo:many}). For example, consider an output label 'V' and timestep 2. Then, input with the label 'V' at the second timestep will be treated as positive if 'V' is present at timestep 2 and negative otherwise, regardless of other labels in other timesteps. After sampling, the algorithm monitors the neurons (Line~\ref{algomany:9} of Algorithm~\ref{algo:many}). Next, it retrieves only observations at the current timestep, as other observations at other timesteps are irrelevant as they can be associated with other output labels.

Next, in \rolled mode, all observations at the currently monitored timestep are concatenated with previous observations at other timesteps as a distinct observation (Line ~\ref{algomany:14} of Algorithm~\ref{algo:many}). For example, for a model with 10 timesteps and 100 examples at each timestep, there will be 1000 observations per neuron in \rolled mode (Line ~\ref{algomany:18} of Algorithm~\ref{algo:many}). However, unlike Algorithm~\ref{algo:one}, it takes 100 observations from timestep 0 when CI is done for timestep 0, ignoring other (9*100) observations. Then, the next 100 observations are taken from timestep 1 when CI is done for timestep 1, ignoring the other 900 observations and so on. Finally, it uses all 1000 observations to identify the concerns (Line ~\ref{algomany:18} of Algorithm~\ref{algo:many}). As such, in many-output models, \textit{rolled}-variant is unaware of the association between an output label and timestep, i.e., contextual information on what labels appear at what step usually. However, in \textit{unrolled} mode, dominant nodes are identified only based on current observations (Line ~\ref{algomany:inittemp} of Algorithm~\ref{algo:many}), and therefore, this variant is capable of identifying concerns in a timestep-wise output-sensitive manner.

\subsection{Tangling Identification}
\label{subsec:ti}
The CI stage mostly identifies nodes relevant to positive samples. However, a module is still required to recognize negative classes. Therefore, tangling identification (TI) aims to bring back some nodes after observing negative samples. TI is particularly important for \relu-based decomposition as the CI stage only keeps the most active nodes after observing positive examples. As a result, it may only recognize the dominant output class, thus becoming a single class classifier. However, for logistic activation-based models, we use a different approach during CI that keeps nodes showing higher central activation tendencies in positive examples. This technique also keeps some tangled nodes as it does not remove nodes that are almost similarly activated in both samples.

For \relu-based models, we bring back a few nodes and edges related to the non-dominant concerns by observing the negative examples. For one-output models, in \textit{rolled} mode, the observations are flattened, similar to the CI stage (Line~\ref{algoone:10} of Algorithm~\ref{algo:one}). Then, it restores a node if it is active in some negative examples ($TI\_relu$ method in Algorithm \ref{algo:cirelu}). In \textit{unrolled} mode, TI is performed in a timestep-sensitive manner as was done for CI. It restores a node in a timestep if it is found to be active in some negative examples in a timestep (Line~\ref{algoone:14} of Algorithm~\ref{algo:one}).

\subsection{Concern Modularization}
\label{subsec:cm}
For Concern Modularization, we channel the output layer to convert $N$ output nodes into two types, dominant ($D$) and non-dominant ($ND$) nodes. For each node in the layer before the output layer, we average all the edges connecting to non-dominant output class nodes. Then, we connect these nodes to the newly introduced non-dominant ones. Next, we remove all other edges from nodes in the preceding layer. Thus, it converts the output layer into a binary classification-based problem. For example, given an input, the module recognizes whether it belongs to a dominant output class. For many-output models, it performs this operation for each timestep, except for the encoder-decoder architecture. For instance, in language translation models, each module produces a single operation translating an input sentence to a different language. 

\section{Evaluation}
\label{sec:evaluation}
This section describes the experimental setup and evaluates our approach using three research questions.

\subsection{Experimental Setups}
\label{subsec:setup}
\subsubsection{Datasets} We perform our experiment on five widely used datasets for text-based sequential problems. Each dataset is used to train different types of RNN models.

\textbf{MathQA~\cite{amini2019mathqa}:} This dataset contains a series of mathematical questions. Each question has a particular tag (e.g., geometry, physics, probability, etc.). Also, the dataset has a total of 6 output classes. 

\textbf{Wiki-toxicity~\cite{wulczyn2017ex}:} This dataset contains Wikipedia pages' comments. Each comment is annotated with seven toxicity labels (e.g., toxic, severe toxicity, obscene, threat, insult, etc.). 

\textbf{Clinc OOS~\cite{larsonetal2019evaluation}:} In NLP, intent classification is a well-known problem in which an input text is categorized based on a user's needs. This dataset has ten output classes.

\textbf{Brown Corpus~\cite{francis1967computational}:} It contains English linguistic data. Each word is annotated with a part-of-speech tag from 12 different tags.

\textbf{Tatoeba~\cite{TIEDEMANN12.463}:} This dataset contains sentences in more than 400 languages. Each sentence in English is translated into other languages. Thus, the dataset is especially used for multilingual evaluation~\cite{ruder-etal-2021-xtreme,artetxe2019massively,fan2021beyond}. We selected English, Italian, German, and French languages as they have the richest (\# of training data) corpus.

\subsubsection{Models} 

For every RNN variant, we built four models for each of the five I/O architectures. Specifically, we used 1, 2, 3, and 4 RNN layers to build models and named them RNN-$<$no. of RNN layers$>$ (RNN refers to either of LSTM, GRU, or Vanilla). 
The structure of the models has been inspired by prior works~\cite{pan2020decomposing}. Moreover, the data pre-processing and architecture of the language models are based on a real-world example~\cite{chollet2021deep}. In model architecture, we use a combination of 8 different Keras layers; (a)~\texttt{Embedding} represents words as a fixed-length high-dimensional vector, (b)~\texttt{RepeatVector} repeats the input n times, (c)~\texttt{Flatten} converts a multidimensional input into a single dimension, (d)~\texttt{SimpleRNN} is the vanilla RNN layer, (e)~\texttt{LSTM}, (f)~\texttt{GRU}, (g)~\texttt{Masking} ignores the padded inputs, and (h)~\texttt{TimeDistributed} applies the same layer across timesteps.

\subsubsection{Evaluation metrics} To evaluate, we use three metrics.

\textbf{Accuracy.} 
For comparing the trained model with the modules, we use testing accuracy as one of the metrics.
We interchangeably use the trained model accuracy (TMA), monolithic model accuracy (MMA), etc. For the modules, we use a voting-based approach (similar to~\cite{pan2020decomposing, pan2022decomposing}) to compute the composed accuracy. All modules in a problem receive the same input, with a joint decision computed at the end. We use the following terminologies interchangeably -- composed model accuracy (CMA) and module accuracy (MA).

\textbf{BLEU Score~\cite{papineni2002bleu}.} For language translations, we use the BLEU score as it is widely applied~\cite{sutskever2014sequence,lin2004rouge}. 

\textbf{Jaccard Index.~\cite{pan2020decomposing}} We compute the Jaccard index (JI) to measure the similarity of the model and the modules.

\subsection{Results}
In this section, we present the results and discuss them briefly. We evaluated the decomposition on 60 models (20 for each RNN variant). Moreover, we repeat the experiments in both \rolled, and \unrolled modes. Due to space limitations, we only present a summary of the results (detailed results can be found here~\cite{rnnrep}). Moreover, for gated RNN variants, we repeat all experiments in two proposed approaches for decomposing gates in \S~\ref{subsec:suplog}. We found that the decomposition cost of both approaches is comparable, and only results from the second approach are presented here. 

\label{subsec:result}
\subsubsection{RQ1: Does decomposing RNN model into modules incur cost? }

% \input{rq1tb}
% \usepackage{multirow}
% \usepackage{graphicx}
\begin{table*}[]
\scriptsize
\centering
{
% \setlength{\extrarowheight}{3pt}%
% \resizebox{0.98\textwidth}{!}
% {%
\setlength\tabcolsep{1.2pt}
\caption{Cost of Decomposing RNN into Modules}

\begin{tabular}{|c|c|cc|c|c|c|c|c|c|c|c|c|c|c|c|c|}
\hline
\textbf{I/O Type}                                & \textbf{Dataset}                & \multicolumn{2}{c|}{\textbf{Mode}}                      & \textbf{LSTM-1} & \textbf{LSTM-2} & \textbf{LSTM-3} & \textbf{LSTM-4} & \textbf{GRU-1}  & \textbf{GRU-2}  & \textbf{GRU-3}  & \textbf{GRU-4}  & \textbf{Vanilla-1} & \textbf{Vanilla-2} & \textbf{Vanilla-3}                      & \textbf{Vanilla-4} & \textbf{Avg. JI} \\ [3pt] \hline \hline
                                                 &                                 & \multicolumn{2}{c|}{Rolled}                             & \textbf{+0.13} & \textbf{+0.07} & -0.03         & \textbf{+0.20} & \textbf{+0.10} & \textbf{+0.10} & \textbf{+0.10} & -0.07         & -0.30            & -0.67            & -0.64                                 & -1.01            & 0.75        \\ \cline{3-17} 
\multirow{-2}{*}{1:1}                     & \multirow{-2}{*}{Math QA}       & \multicolumn{2}{c|}{Unrolled}                           & {\cellcolor[rgb]{ 0,  0,  0}\textbf{N/A}}             & {\cellcolor[rgb]{ 0,  0,  0}\textbf{N/A}}             & {\cellcolor[rgb]{ 0,  0,  0}\textbf{N/A}}             & {\cellcolor[rgb]{ 0,  0,  0}\textbf{N/A}}             & {\cellcolor[rgb]{ 0,  0,  0}\textbf{N/A}}             & {\cellcolor[rgb]{ 0,  0,  0}\textbf{N/A}}             & {\cellcolor[rgb]{ 0,  0,  0}\textbf{N/A}}             & {\cellcolor[rgb]{ 0,  0,  0}\textbf{N/A}}             & {\cellcolor[rgb]{ 0,  0,  0}\textbf{N/A}}                & {\cellcolor[rgb]{ 0,  0,  0}\textbf{N/A}}                & {\cellcolor[rgb]{ 0,  0,  0}\textbf{N/A}}                                     & {\cellcolor[rgb]{ 0,  0,  0}\textbf{N/A}}                & {\cellcolor[rgb]{ 0,  0,  0}\textbf{N/A}}         \\ \hline
                                                 &                                 & \multicolumn{2}{c|}{Rolled}                             & -0.13         & -0.44         & -1.40         & -0.44         & -0.20         & \textbf{+0.22} & -0.22         & -0.47         & -1.58            & -5.51            & -2.27                                 & -2.40            & 0.85        \\ \cline{3-17} 
\multirow{-2}{*}{M:1}                    & \multirow{-2}{*}{\makecell{Clinc \\ OOS}}  & \multicolumn{2}{c|}{Unrolled}                           & \textbf{+0.96} & \textbf{+0.51} & \textbf{+0.07} & -0.67         & \textbf{+0.09} & \textbf{+0.00} & -0.09         & -0.69         & -0.07            & -1.38            & -0.87                                 & -0.11            & 0.86        \\ \hline
                                                 &                                 & \multicolumn{2}{c|}{Rolled}                             & -14.73        & -14.98        & -94.59        & -80.78        & -0.49         & -1.82         & -6.05         & -1.02         & -15.21           & -14.96           & -15.15                                & -15.28           & 0.86        \\ \cline{3-17} 
\multirow{-2}{*}{1:N}                    & \multirow{-2}{*}{\makecell{Toxic \\ Comment}} & \multicolumn{2}{c|}{Unrolled}                           & -0.01         & \textbf{+0.05} & \textbf{+0.17} & \textbf{+0.44} & -0.27         & -0.76         & \textbf{+0.04} & -0.14         & -0.97            & -1.48            & -0.36                                 & -2.78            & 0.86        \\ \hline
                                                 &                                 & \multicolumn{2}{c|}{Rolled}                             & -77.07        & -76.56        & -80.55        & -84.63        & -75.28        & -75.48        & -78.04        & -78.11        & -80.13           & -80.18           & -79.93                                & -79.93           & 0.83        \\ \cline{3-17} 
\multirow{-2}{*}{M:N}                   & \multirow{-2}{*}{\makecell{Brown \\ Corpus}}  & \multicolumn{2}{c|}{Unrolled}                           & -0.52         & -0.22         & -2.04         & -2.84         & -0.02         & -0.77         & -0.52         & -2.59         & -1.02            & -1.64            & -2.90                                 & -3.37            & 0.83        \\ \hline
                                                 &                                 & \multicolumn{1}{c|}{}                           & EN-FR & \textbf{+0.02} & \textbf{+0.18} & -0.45         & -0.49         & \textbf{+0.25} & -0.06         & \textbf{+0.02} & -0.05         & \textbf{+0.13}    & \textbf{+0.16}    & -0.68                                 & -1.04            & 0.80        \\ \cline{4-17} 
                                                 &                                 & \multicolumn{1}{c|}{}                           & EN-DE & \textbf{+0.05} & \textbf{+0.06} & -0.07         & -0.15         & -0.16         & \textbf{+0.03} & \textbf{+0.01} & \textbf{+2.01} & -0.02            & -0.48            & -0.56                                 & -0.28            & 0.80        \\ \cline{4-17} 
                                                 &                                 & \multicolumn{1}{c|}{\multirow{-3}{*}{Rolled}}   & EN-IT & \textbf{+0.00} & \textbf{+0.22} & \textbf{+0.08} & -0.18         & \textbf{+0.44} & \textbf{+0.45} & \textbf{+0.99} & \textbf{+0.33} & -0.57            & -1.60            & \textbf{+2.38}                         & \textbf{+2.51}    & 0.80        \\ \cline{3-17} 
                                                 &                                 & \multicolumn{1}{c|}{}                           & EN-FR & \textbf{+0.04} & \textbf{+0.31} & -0.18         & \textbf{+0.14} & \textbf{+0.44} & -0.45         & -0.50         & -0.11         & -0.25            & \textbf{+0.25}    & -0.38                                 & -2.46            & 0.79        \\ \cline{4-17} 
                                                 &                                 & \multicolumn{1}{c|}{}                           & EN-DE & \textbf{+0.01} & \textbf{+0.00} & -0.35         & -0.35         & \textbf{+0.15} & -0.17         & -0.65         & -0.02         & -0.85            & -0.26            & -3.00                                 & -0.34            & 0.79        \\ \cline{4-17} 
\multirow{-6}{*}{\makecell{M:N\\ (Encoder\\-\\Decoder)}} & \multirow{-6}{*}{Tatoeba}       & \multicolumn{1}{c|}{\multirow{-3}{*}{Unrolled}} & EN-IT & -0.10         & \textbf{+0.12} & -0.03         & -0.25         & -0.51         & -0.26         & -0.26         & -0.50         & \textbf{+0.76}    & -0.52            & \textbf{+3.59} & -0.48            & 0.79        \\ \hline
\end{tabular}%
\label{tab:rq1}
\\All values are in \% except Avg. JI. Here, 1:1=one-to-one,M:1=many-to-one,1:N=one-to-many,M:N=many-to-many
}
% }
\end{table*}

In this research question, we evaluate the cost of decomposition in 60 scenarios. To that end, we determine the quality of the decomposition and composition approaches. First, we decompose a trained RNN model into modules. Each module receives the same input and recognizes an output class. Next, we use the modules to compose a new model using a voting-based approach. The modules' decisions are combined into one that matches the output type. For example, the final decision is a single output class for \{one, many\}-to-one. Whereas, for \{one, many\}-to-many, the final decision will be a list of output classes. Then we compare the composed accuracy with the monolithic one.

We apply the \rolled and \unrolled-variants to decompose the 60 models, and the decomposition cost is depicted in Table~\ref{tab:rq1} in terms of accuracy difference, $\delta=CMA-MMA$. In the \rolled-variant, we identify the concern for all the timesteps at once. For the \unrolled-variant, we check the concerns for each timestep separately after unrolling loops in the RNN model. We evaluate the \unrolled-variant with the \textit{M:1}, \textit{1:N}, and \textit{M:N} architectures. The unrolled-variant does not apply to the \textit{1:1} architecture as it does not contain loops. We found an average loss of 25.8\% (median: -2.04\%) accuracy for the \rolled-variant. In contrast, the average accuracy loss is 0.74\% (median: -0.44\%) for the \unrolled one. Moreover, in 31.25\% scenarios, CMA remains the same or improves (considering \rolled mode for \textit{1:1} and \unrolled for others). 

For the \rolled-variant, the bulk of the accuracy losses come from the many-output models, while the differences in the one-output model are trivial (Table~\ref{tab:rq1}). In \rolled mode, the average accuracy loss for one-output models is -0.7\%, while it is -50.87\% for many-output models. However, in \unrolled mode, decomposition quality for both one and many-output models are comparable (avg. loss for one-output: -0.2\% and for many-output: -1.02\%). This is because \rolled-variant is insensitive to timestep, therefore, more likely to lose output-related contextual information at a timestep. In other words, timestep-specific output sensitivity is lost in \rolled-variant. Many-output models, where each timestep has an output, require that concern is identified for that output based on what is observed in that timestep, which \unrolled-variant does. 
On the other hand, one-output models only rely on the hidden state of the final \rnn cell, requiring no output sensitivity for different timesteps. Therefore, we recommend \unrolled-variant for many-output models, particularly where each timestep-output is subject to decomposition.

For language models, we compute the BLEU score for each pair of languages. This score measures their translation quality~\cite{papineni2002bleu}. We found an average gain of 0.10\% (median: +0.01\%) for the \rolled-variant in the BLEU score. While, in \unrolled mode, the average BLEU score loss is -0.2\% (median: -0.25\%). Based on the argument by~\cite{bleu}, such a change in the BLEU score does not affect the quality of the translation. Furthermore, for 52.8\% cases in \rolled mode, the composed model's BLEU score remains the same or improves compared to the original one.

\textbf{Similarity:} Apart from the cost, we also measure the structural similarity (in terms of learned parameters) between the monolithic model and module to assess the effectiveness of decomposition. A high similarity indicates an ineffective decomposition approach, creating modules replicating the original model. To evaluate the variability among modules and models, we compute Jaccard Index (JI). We found that, on average, the \textit{JI} value for the rolled-based approach is 0.82, and for the unrolled one is 0.83. This result shows that the modules are significantly different from the parent models. Overall, we found that the RNN model can be decomposed into modules at a very small cost. Also, the decomposed modules are significantly different from the original model.

\subsubsection{RQ2: Can Decomposed Modules be Reused to Create a New Problem?}
In this RQ, we reuse the decomposed modules. Our evaluation focuses on two reusability cases: a) (re)use the modules within the same input-output (I/O) type, and b) from a different type, to create a new problem. 
We perform these experiments separately for three RNN variants (LSTM, GRU, and Vanilla). Next, we discuss the results from each case.

 \begin{table}[http]
\vspace{5pt}
\scriptsize
\centering
{%
\setlength\tabcolsep{1.2pt}
\caption{Summary of intra and inter-reuse experiments}
\begin{tabular}{|c|c|cc|cc|cc|}
\hline
\multirow{2}{*}{\textbf{Reuse Type}} & \multirow{2}{*}{\textbf{I/O Type}} & \multicolumn{2}{c|}{\textbf{LSTM}}                     & \multicolumn{2}{c|}{\textbf{GRU}}                      & \multicolumn{2}{c|}{\textbf{Vanilla}}                      \\ \cline{3-8} 
                                     &                                      & \multicolumn{1}{c|}{Mean}            & Median          & \multicolumn{1}{c|}{Mean}            & Median          & \multicolumn{1}{c|}{Mean}            & Median          \\ \hline
\multirow{5}{*}{Intra}               & 1:1                                  & \multicolumn{1}{c|}{-0.07}         &  \textbf{+0.00} & \multicolumn{1}{c|}{-0.03}         &  \textbf{+0.00} & \multicolumn{1}{c|}{-0.05}         & \textbf{+0.00} \\ \cline{2-8} 
                                     & M:1                                  & \multicolumn{1}{c|}{-0.82}         & -0.67         & \multicolumn{1}{c|}{-0.73}         & -0.44         & \multicolumn{1}{c|}{-0.53}         & -0.44         \\ \cline{2-8} 
                                     & 1:N                                  & \multicolumn{1}{c|}{-2.20}         & -0.31         & \multicolumn{1}{c|}{-5.49}         & -1.25         & \multicolumn{1}{c|} {\textbf{+0.38}} &  \textbf{+0.02} \\ \cline{2-8} 
                                     & M:N                                  & \multicolumn{1}{c|}{-0.50}         & -0.71         & \multicolumn{1}{c|}{\textbf{+0.42}}&  \textbf{+0.24} & \multicolumn{1}{c|}{ \textbf{+0.01}} & -0.02         \\ \cline{2-8} 
                                     & \makecell{M:N\\ Encoder-Decoder }               & \multicolumn{1}{c|}{ \textbf{+5.20}} &  \textbf{+3.49} & \multicolumn{1}{c|}{ \textbf{+3.83}} & -1.31         & \multicolumn{1}{c|}{ \textbf{+4.17}} &  \textbf{+3.00} \\ \hline
\multirow{2}{*}{Inter}               & 1:1-1:N                              & \multicolumn{1}{c|}{-2.49}         & -1.25         & \multicolumn{1}{c|}{-7.93}         & -3.10         & \multicolumn{1}{c|}{-8.23}         & -7.68         \\ \cline{2-8} 
                                     & M:1-M:N                              & \multicolumn{1}{c|}{-2.92}         & -1.68         & \multicolumn{1}{c|}{-3.28}         & -2.50         & \multicolumn{1}{c|}{-3.55}         & -2.63         \\ \hline
\end{tabular}%
\label{tab:rq2}
\\All values are in \%.
}
\end{table}

\textbf{Intra RNN Type Reuse.} To evaluate this reuse type, we take modules from the same I/O type of RNN. To do so, we use the dataset available in our benchmark. We take two modules, compose them, and evaluate the accuracy of the composed models. 
Additionally, we train a model with the same model architecture of the modules with examples of the dominant classes of the modules. For example, consider the case of intra-reuse for a \textit{M:1} \textit{GRU} model trained on the Clinc OOS dataset. In this case, consider the two output toxicity levels: \textit{severe} and \textit{threat}. First, a model is trained from scratch using inputs of these two labels alone to reuse them. Then, we take corresponding modules from a previously decomposed model with all labels and compare their composed accuracy with that of the trained one newly.

We take 2 modules from each trained model and build a sub-problem. The total possible combinations for taking 2 modules from a model trained with a dataset having $N$ output classes are $N \choose 2$. Our benchmark has 6, 7, 10, and 12 output classes in the datasets used to train the models for \textit{1:1}, \textit{1:N}, \textit{M:1}, and \textit{M:N} (traditional) architectures, respectively. So, the total possible combinations can be 152 ($6 \choose 2$ + $7 \choose 2$ + $10 \choose 2$ + $12 \choose 2$). For each case, we train a model from scratch using two labels. Then, to get composed accuracy from reused modules, we consider the modules decomposed from RNN-4 (one with the highest number of layers), similar to prior work~\cite{pan2020decomposing}. Then, we compare their accuracies to understand the effectiveness of intra-reuse. We found that for \textit{1:1},  \textit{1:N}, \textit{M:1}, and \textit{M:N} (traditional), the change of accuracy is -0.05\% (median 0\%), -2.44\% (median -0.17\%), -0.69\% (median -0.56\%), and -0.02\% (median 0\%), respectively.

In Table~\ref{tab:rq2}, we report the results. Overall, there is a slight loss (mean: -0.58\%, median: -0.11\%) of accuracy when modules are reused. We also perform a similar evaluation for language translation modules. Since there are 3 modules produced from the trained model, taking 2 modules at a time can create 3 possible scenarios. We also train a multi-lingual model that takes English sentences as input and converts it into the 2 chosen non-English languages. We report the results in Table~\ref{tab:rq2}. We found that there is an average gain of 4.40\% BLEU score (median: +2.66\%) for each language pair.

\textbf{Inter RNN Type Reuse.} We use modules from different I/O architecture types to evaluate this reuse type to build a new problem. However, to compose the modules, all of them should be able to process similar input types. For instance, assume taking a module from an RNN model that receives one input (1-to-\{1, N\}) and reusing it with a module from a model that receives several inputs at a time (M-to-\{1, N\}). This composition would not work because the two modules do not satisfy the same input constraints. For that, we evaluate in two different settings. First, \textbf{1-to-\{1, N\}}, in which we take one module decomposed from a \textit{1:1} model and another from a \textit{1:N} model. Then we compose them together to form a model. Second, \textbf{M-to-\{1, N\}}, in which we take one module decomposed from the \textit{M:1} model and another from \textit{M:N} (traditional) model. The encoder-decoder architecture prevents model reuse with other I/O types for language-translation models. However, such modules can be reused if decomposed from different datasets. We discuss such an experiment in \S\ref{subsubsec:motivate}, when we recreate the motivation scenario and show the possibilities of solving the problem.
We found that for the former scenario (1-to-\{1, N\}), there is an -6.22\% (median -2.95\%) loss of accuracy, on average. Whereas, for the latter scenario (M-to-\{1, N\}), the loss is, on average, -3.25\% (median -2.20\%).

\subsubsection{RQ3: Can Decomposed Modules be Replaced?}
This RQ investigates how to replace a decomposed module with another one. Similar to RQ2, we evaluate two different scenarios--(a) replacing a module with another performing the same operation within the same I/O type and (b) between different I/O types (in our case, different datasets too). We perform these experiments for different RNN-variants separately. We discuss each scenario in the following paragraphs.

\begin{table}[]
\scriptsize
\centering
% {\setlength{\extrarowheight}{3pt}
% \resizebox{0.45\textwidth}{!}{%
{
\setlength\tabcolsep{1.2pt}
\caption{Summary of intra and inter-replace experiments}
\begin{tabular}{|c|c|cc|cc|cc|}
\hline
\multirow{2}{*}{\textbf{Replace Type}} & \multirow{2}{*}{\textbf{I/O Type}} & \multicolumn{2}{c|}{\textbf{LSTM}}                     & \multicolumn{2}{c|}{\textbf{GRU}}                      & \multicolumn{2}{c|}{\textbf{Vanilla}}                      \\ \cline{3-8} 
                                       &                                    & \multicolumn{1}{c|}{Mean}            & Median          & \multicolumn{1}{c|}{Mean}            & Median          & \multicolumn{1}{c|}{Mean}            & Median          \\ \hline
\multirow{5}{*}{Intra}                 & 1:1                                & \multicolumn{1}{c|}{\textbf{+0.52}} & -0.49         & \multicolumn{1}{c|}{\textbf{+0.26}} & \textbf{+0.25} & \multicolumn{1}{c|}{-0.55}         & -0.54         \\ \cline{2-8} 
                                       & M:1                                & \multicolumn{1}{c|}{-0.30}         & -0.13         & \multicolumn{1}{c|}{-0.20}         & -0.14         & \multicolumn{1}{c|}{-2.57}         & -1.52         \\ \cline{2-8} 
                                       & 1:N                                & \multicolumn{1}{c|}{-0.74}         & \textbf{+0.04} & \multicolumn{1}{c|}{-0.07}         & -0.24         & \multicolumn{1}{c|}{-2.39}         & -2.48         \\ \cline{2-8} 
                                       & M:N                                & \multicolumn{1}{c|}{-1.41}         & -1.56         & \multicolumn{1}{c|}{-1.81}         & -2.90         & \multicolumn{1}{c|}{-4.76}         & -4.04         \\ \cline{2-8} 
                                       & \makecell{M:N\\ Encoder-Decoder }               & \multicolumn{1}{c|}{\textbf{+0.30}} & \textbf{+0.16} & \multicolumn{1}{c|}{\textbf{+1.72}} & \textbf{+1.88} & \multicolumn{1}{c|}{\textbf{+0.92}} & \textbf{+0.96} \\ \hline
\multirow{2}{*}{Inter}                 & 1:1-1:N                            & \multicolumn{1}{c|}{-4.75}         & -0.02         & \multicolumn{1}{c|}{-5.10}         & -3.30         & \multicolumn{1}{c|}{-4.08}         & -6.60         \\ \cline{2-8} 
                                       & M:1-M:N                            & \multicolumn{1}{c|}{-10.48}        & -12.23        & \multicolumn{1}{c|}{-6.71}         & -7.11         & \multicolumn{1}{c|}{-12.23}        & -13.68        \\ \hline
\end{tabular}%
\label{tab:rq3}
\\All values are in \%. 
% @Astha make positive values as +then value. Also, color the cells instead of just bold.
}
% }
\end{table}

\textbf{Intra RNN Type Replacement.} Here, a module is replaced with another one performing the same operation. The rationale of this experiment is to investigate how decomposition can help fix faulty models (e.g., low accuracy). To that end, we take the model with the lowest performance score. Then, we replace a module with one decomposed from the model with the highest accuracy in that category. For instance, consider the case of intra-replace for \textit{1:1} \textit{LSTM} models trained on the Math QA dataset. In this case, LSTM-1 performs best and worst for LSTM-4. Therefore, we replace a module from LSTM-4 with one decomposed from LSTM-1. Then, we compute the accuracy of the composed model. Table~\ref{tab:rq3} shows the result of the experiments for all types of RNN models. We found that for a model trained with (\textit{1:1}, Math QA), (\textit{M:1}, Clinc OOS), (\textit{1:N}, Toxic comment), and (\textit{M:N}, Brown corpus) architecture-dataset pair, the average change of accuracy is +0.07\% (median +0.25\%), -1.03\% (median -0.42\%), -1.07\% (median -0.27\%), and -2.66\% (median -2.90\%), respectively. For the \textit{M:N} encoder-decoder architecture, we replace the modules from the lowest average BLEU score model with the highest. As a result, we observed a 0.98\% increase in the BLEU score compared to the monolithic model's BLEU score.

\textbf{Inter RNN Type Replacement.} Here, we replace a module with another one between different I/O types. We use the resulting composed model to perform different tasks. For instance, we replace a module from a model with a \textit{1:1} architecture with one from a \textit{1:N} model. Similar to the RQ2, the replaced module must accept a similar input type. For this reason, we perform two different experiments. First, for \textbf{1-to-\{1, N\}} architectures, we replace a module decomposed from \textit{1:1} with one from an RNN model using the \textit{1:N} architecture. Second, for the \textbf{M-to-\{1, N\}}, we replace a module decomposed from a \textit{M:1} with a module from an RNN model with \textit{M:N} architecture (traditional). Like RQ2, for \textit{M:N} encoder-decoder architecture, we cannot perform the inter-RNN type replaceability due to the difference in the I/O architecture.

We found that for the 1-to-\{1, N\} replaceability, there is an average -4.64\% (median -3.85\%) accuracy loss. Whereas for M-to-\{1, N\} replaceability, the loss is -9.81\% (median -11.49\%) (Table~\ref{tab:rq3}). Overall, the loss is -8.47\% (median -10.79\%) for the inter-model type replaceability.

\subsubsection{Recreating Motivating Examples}
\label{subsubsec:motivate}
Here, we evaluate the scenarios discussed in the \S\ref{sec:motivation}. For the first use case, a new language needs to be added to an existing model. We created a model with the languages from the motivating example (i.e., English, French, German, and Italian). Then, we decompose it to create modules. We train another model with the Ukrainian language as one of the target languages and decompose it too. Next, we compose the modules from the original model with the module that translates English to Ukrainian. In the second approach, we train a new model that translates only English to Ukrainian and uses it as a module. Lastly, we compose it with the previously decomposed modules. We found that both approaches can address the problem. However, the average BLEU score of modules is slightly less than that of the monolithic model for the first approach (see Table~\ref{tb:motivating}). Here, we only report the results for LSTM models. Results for other models are similar and included in the replication package~\cite{rnnrep}.

For the second scenario, we replace the module from a model that performs badly with a module decomposed from a model that performs better. We build a model that translates the Estonian language into English, Italian, and German. We decompose the model into modules and replace the Estonian with the English module with two approaches described in the examples. We found that both approaches perform better than the trained model from scratch.

% Table generated by Excel2LaTeX from sheet 'Sheet2'
\begin{table}[htbp]
% \setlength{\parskip}{.1cm}
% %	\setlength{\intextsep}{.1cm plus .1cm minus 1.cm}
\setlength{\belowcaptionskip}{.01cm}
  \centering
%   \vspace{5pt}
  \scriptsize
  \setlength\tabcolsep{1pt}
  
  \caption{Motivating Scenarios (Results for LSTM models)}
%   \vspace{-5pt}
    \begin{tabular}{|l|r|r|r|}
    \hline
    \multicolumn{1}{|c|}{\textbf{Scenario}} & \multicolumn{1}{c|}{\textbf{TMA}} & \multicolumn{1}{c|}{\textbf{MA1}} & \multicolumn{1}{c|}{\textbf{MA2}} \\
    \hline
    \hline
    Add Ukranian Language &  32.12\%     &  31.94\%    &  \textbf{32.53\%}\\
    \hline
    Update Estonian-English Translation &  20.80\%     &  \textbf{20.89\%}     & \textbf{21.30\%} \\
    \hline
    \end{tabular}%
  \label{tb:motivating}\\
  \vspace{1pt}
  * MA\{X\}, TMA: Avg. BLEU score for scenario X and trained model.
\end{table}%

\paragraph{Summary} We found that decomposing trained RNN models into modules has a trivial cost (accuracy: -0.6\% and BLEU score: +0.10\%). Also, these decomposed modules can be reused (accuracy: -2.38\%, BLEU: +4.40\%) and replaced (accuracy: -7.16\%, BLEU: +0.98\%) in various scenarios.

\section{Threats to validity}
\label{sec:threat}

\textbf{Internal threat:}
An internal threat can be the trained models. To mitigate, we follow prior works~\cite{pan2020decomposing,pan2022decomposing,sutskever2014sequence} to build the model (details in \S-V.A).
Another threat can arise from the stochastic nature of DL. To mitigate, in RQ1, each task is evaluated on four different model architectures, and in RQ2 and RQ3, every combination is exhaustively evaluated.

\textbf{External threat:}
An external threat can be the experimental datasets. To mitigate this, we chose canonical datasets already used in the literature~\cite{amini2019mathqa,francis1967computational,wulczyn2017ex,larsonetal2019evaluation,chollet2021deep}. These datasets have a rich corpus and are adequately diverse to allow evaluation of our technique in different practical usage of NLP, i.e., single-output, multi-output, and generation (language translation).
\section{Conclusion and future directions}
\label{sec:conclusion}

Modularization and decomposition have been shown to enable many benefits in traditional software, such as reuse, replacement, hiding changes, and increased comprehensibility of the modules~\cite{parnas1972criteria,dijkstra1970notes}. Recent works have demonstrated that DL systems can also benefit from such a decomposition and demonstrate these advantages for FCNN and CNN networks~\cite{pan2020decomposing,pan2022decomposing}. This paper further advances our knowledge of modularity in the context of DL systems by extending it to RNNs, an important class of DNNs. It shows that different RNN models can be effectively decomposed and reused in different scenarios. Practitioners can use modules to compose new models. Also, they can leverage modules to replace faulty parts of existing models. The approach has been evaluated extensively on a benchmark of 60 models in different setups, i.e., different input/output types, RNN variants, and assuming both non-linear and logistic activation functions, etc. We found that decomposition has a small cost in terms of performance (accuracy and BLEU score). While this work limits its focus on the reuse and replace dimension of the modularity, we envision this decomposition can also enable/facilitate other benefits such as:

\textit{Hiding Changes:} One of the key benefits of modularization is its ability to isolate and hide changes to a smaller number of components. This notion could be extended to deep learning software, making maintenance of large models, particularly in NLP, more manageable. Given the significance of change hiding in these scenarios, it is worth exploring the potential of the proposed modularization to streamline the maintenance process.

\textit{Increase Comprehensibility:} Modularization has been shown to enhance our understanding of program logic, as noted by Dijkstra~\cite{dijkstra1970notes}. In the context of deep learning models, modularization could help reveal the internal logic more efficiently by breaking down a monolithic black-box model into distinct, functional units.

\section{Acknowledgement}
This work was supported in part by US NSF grants CNS-21-20448
and CCF-19-34884. We want to thank the reviewers for their valuable and insightful comments. The views expressed in this work are solely those of the authors and do not reflect the opinions of the sponsors.

\balance
\bibliographystyle{IEEEtranN}
\bibliography{refs}

%\input{appendix/appendix.tex}

\end{document}